\documentclass{article}

\usepackage{arxiv}
\usepackage{amssymb}
\usepackage{amsmath}
\usepackage{empheq}
\usepackage{amscd}
\usepackage{graphicx}
\usepackage{graphics}
\usepackage[noadjust]{cite}
\usepackage{amsthm}
\usepackage{caption}
\usepackage{multirow}
\usepackage{array}
\usepackage{rotating}

\usepackage{indentfirst}
\usepackage{tikz}
\usepackage{calc}
\usepackage{epsfig}
\usepackage{natbib}
\usepackage[makeroom]{cancel}
\usepackage{multicol}
\usepackage{csquotes}
\usepackage{algorithm}
\usepackage{algorithmic}
\usepackage{enumitem}
\usepackage{hyperref}
\usepackage{url}
\hypersetup{colorlinks,linkcolor={blue},citecolor={blue},urlcolor={blue}}

\usepackage{timet}
\usepackage{graphicx}
\usepackage{multirow}
\usepackage{multicol}
\usepackage{lipsum}
\usepackage{timet}
\usepackage{epsfig}
\usepackage{amsmath}
\usepackage{amsfonts}
\usepackage{amssymb}
\usepackage{color}
\numberwithin{equation}{section}
\usepackage{caption}
\usepackage{float}
\usepackage{subcaption}
\usepackage{graphics}
\usepackage{xcolor,graphicx}
\usepackage{csquotes}
\usepackage{mathtools}
\usepackage{optidef}

\newcommand{\vertiii}[1]{{\left\vert\kern-0.25ex\left\vert\kern-0.25ex\left\vert #1 
    \right\vert\kern-0.25ex\right\vert\kern-0.25ex\right\vert}}

\usepackage{hyperref}       

\begin{document}
\title{Accurate 3D frequency-domain seismic wave modeling with the wavelength-adaptive 27-point finite-difference stencil: a tool for full waveform inversion}

\author{\href{http://orcid.org/0000-0003-1805-1132}{\includegraphics[scale=0.06]{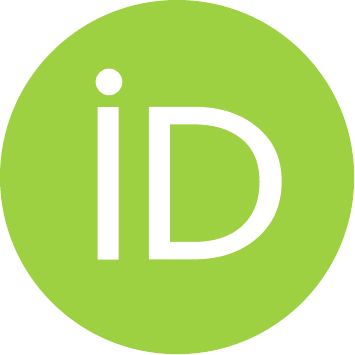}\hspace{1mm}Hossein S. Aghamiry} \\
  University Cote d'Azur - CNRS - IRD - OCA, Geoazur, Valbonne, France. 
  \texttt{aghamiry@geoazur.unice.fr}
\And
  \href{https://orcid.org/0000-0002-9879-2944}{\includegraphics[scale=0.06]{orcid.pdf}\hspace{1mm}Ali Gholami} \\
  Institute of Geophysics, University of Tehran, Tehran, Iran.
  \texttt{agholami@ut.ac.ir} \\ 
  \And
\href{http://orcid.org/0000-0003-2775-7647}{\includegraphics[scale=0.06]{orcid.pdf}\hspace{1mm}Laure Combe} \\ 
  University Cote d'Azur - CNRS - IRD - OCA, Geoazur, Valbonne, France. 
  \texttt{combe@geoazur.unice.fr} 
\And
\href{http://orcid.org/0000-0002-4981-4967}{\includegraphics[scale=0.06]{orcid.pdf}\hspace{1mm}St\'ephane Operto} \\ 
  University Cote d'Azur - CNRS - IRD - OCA, Geoazur, Valbonne, France. 
  \texttt{operto@geoazur.unice.fr} 
 }

\renewcommand{\shorttitle}{Accurate 3D FDFD, Aghamiry et al.}

\maketitle

\begin{abstract}
Efficient frequency-domain Full Waveform Inversion (FWI) of long-offset/wide-azimuth node data can be designed with a few discrete frequencies. However, 3D frequency-domain seismic modeling remains challenging since it requires solving a large and sparse linear indefinite system per frequency. When such systems are solved with direct methods or hybrid direct/iterative solvers, based upon domain decomposition preconditioner, finite-difference stencils on regular Cartesian grids should be designed to conciliate compactness and accuracy, the former being necessary to mitigate the fill-in induced by the Lower-Upper (LU) factorization.
Compactness is classically implemented by combining several second-order accurate stencils covering the eight cells surrounding the collocation point, leading to the so-called 27-point stencil. Accuracy is obtained by applying optimal weights on the different stiffness and consistent mass matrices such that numerical dispersion is jointly minimized for several number of grid points per wavelength ($G$). However, with this approach, the same weights are used at each collocation point, leading to suboptimal accuracy in heterogeneous media. In this study, we propose a straightforward recipe to improve the accuracy of the 27-point stencil. First, we finely tabulate the values of $G$ covering the range of wavelengths spanned by the subsurface model and the frequency. Then, we estimate with a classical dispersion analysis in homogeneous media the corresponding table of optimal weights that minimize dispersion for each $G$ treated separately. We however apply a Tikhonov regularization to guarantee smooth variation of the weights with $G$. Finally, we build the impedance matrix by selecting the optimal weights at each collocation point according to the local wavelength, hence leading to a wavelength-adaptive stencil.
We validate our method against analytical solutions in homogeneous and velocity gradient models and 3D heterogeneous benchmarks with sharp contrasts and complex tectonic pattern. In the latter case, the accuracy of the adaptive stencil is checked with the highly accurate solution of the convergent Born series (CBS).  Each benchmark reveals the higher accuracy of the adaptive stencil relative to the non-adaptive one. We also show that, in the presence of sharp contrasts, the adaptive stencil is more accurate than a $\mathcal{O}(\Delta t^2, \Delta h^8)$ finite-different time-domain method.
\end{abstract}

\section{Introduction}
Numerical seismic wave simulation as the forward problem of Full Waveform Inversion (FWI) can be implemented in the time or frequency domains \citep{Plessix_2006_RTD,Plessix_2007_HIS,Vigh_2008_CWI,Operto_2015_ETF,Plessix_2017_CAT,Kostin_2019_DFA}. Although the time-domain formulation implemented with explicit (matrix-free) time-stepping schemes has emerged as the way forward because forward engines of reverse time migration can be readily used for FWI, frequency-domain FWI still has some potential key strengths when applied on wide-azimuth long-offset stationary-recording acquisitions such as seabed acquisitions implemented with ocean bottom cables or nodes and land acquisitions. These strengths mostly rely on the fact that reliable FWI can be performed with a few discrete frequencies by reducing the redundancy with which such acquisitions locally sample subsurface wavenumbers \citep{Pratt_1990_ITA,Pratt_1999_SWIb,Sirgue_2004_EWI}. This frequency decimation potentially leads to computationally efficient algorithms and a compact volume of data. 
Moreover, the frequency domain provides the most convenient framework to design frequency continuation strategies, which are useful to mitigate the FWI nonlinearity generated by cycle skipping \citep[e.g.,][]{Gorszczyk_2017_TRW}. 
Finally, implementation of attenuation is straightforward and does not generate computational overheads \citep{Toksoz_1981_GRS}. 

The main challenge in frequency-domain FWI is the solution of the time-harmonic wave equation, a boundary-value problem that requires the solution of large and sparse indefinite unsymmetric linear system per frequency with multiple right-hand sides. Although frequency-domain FWI can also be implemented with a time-domain forward engine, when the monochromatic wavefields are extracted on the fly in the loop over time steps by phase sensitive detection \citep{Nihei_2007_FRM} or discrete Fourier transform \citep{Sirgue_2008_FDW}, the benefit of the straightforward implementation of attenuation is lost with this approach. Two main categories of linear algebra methods exist to solve the linear system resulting from the discretization of the time-harmonic wave equation: Direct methods first perform a Lower-Upper (LU) decomposition of the impedance matrix before computing the solutions by forward/backward elimination \citep{Duff_2017_DSS}.  
The strength of this approach is predictability (the solution is obtained in a finite number of operations) and the efficiency of the solution step when a large number of right-hand sides (namely, sources) should be processed as in seismic imaging. The drawbacks are the memory overhead induced by the storage of the LU factors and the limited scalability of the LU decomposition. This approach is generally considered as intractable for 3D applications although recent developments of sparse direct solvers \citep{Wang_2011_OMS,Amestoy_2015_IMM,Amestoy_2018_CBL,Kostin_2019_DFA} and FWI case studies \citep{Operto_2015_ETF,Amestoy_2016_FFF,Operto_2018_MFF} proved precisely the opposite. 
The second category of linear algebra methods relies on iterative solvers \citep{Saad_2003_IMS}, which are scalable but raise the challenge of designing a suitable preconditioner for indefinite Helmholtz problems \citep{Gander_2012_HEL}. Domain decomposition preconditioners based on additive Schwarz are among the most popular ones to tackle this challenge \citep{Dolean_2015_IDD}, where a Krylov subspace iterative method such as the generalized minimal residual (GMRES) algorithm \citep{Saad_1986_GMRES} is used to solve the preconditioned system while direct methods are used to solve the local problems associated with the subdomains of the computational mesh \citep{Tournier_2019_MTI}. 

In both approaches, the discretization scheme should be designed with care to find the best trade-off between the number of degrees of freedom in the computational mesh, accuracy, and compactness. In the frequency-domain, the second-order formulation of the wave equation is generally used to minimize the number of degrees of freedom at the expense of a first-order velocity-stress formulation, which is more popular in the time domain \citep{Virieux_1986_PSW}. Compactness is useful to minimize the numerical bandwidth of the matrix and hence mitigate the memory burden induced by the LU factorization of the global or local matrices as well as the amount of inter-domain communication when parallel domain-decomposition preconditioner are used. Two- and three-dimensional finite-difference stencils on regular Cartesian grid (constant grid interval) have been designed in the past to fulfil these specifications \citep{Jo_1996_OPF,Min_2000_IFD,Hustedt_2004_MGS,Stekl_1998_AVM,Operto_2007_FDFD,Chen_2012_DMF,Turkel_2013_CSO,Chen_2013_OFD,Operto_2014_FAT,Gosselin_2014_FDF}. \citet{Turkel_2013_CSO} develop a 3D sixth-order accurate scheme 27-point finite-difference stencil for the Helmholtz equation with variable wavenumber. In this approach derivatives of the Helmholtz equation are used to eliminate high order derivatives in the discretization error. One difficulty with this approach, regardless its technical complexity, is related to the right-hand side of the discretized equation, which involve high-order derivatives of the physical source \citep[][ Their equation 27]{Turkel_2013_CSO}. This might raise some accuracy issues when the point sources in seismic application doesn't coincide with a grid point \citep{Hicks_2002_ASR}. Essentially, it seems that this approach still lacks validation against highly heterogeneous and contrasted media. The other above mentioned references rely on a different paradigm for compact discretization of the Helmholtz equation, which is more oriented toward numerical optimization. Compactness is achieved with second-order accurate stencils while accuracy is achieved by linearly combining several second-order accurate stencils built on different (rotated) coordinate systems with appropriate weights and by designing a consistent mass matrix as opposed to the lumped mass matrix.  The weights used to combine the different stiffness matrices and the different entries of the consistent mass matrix are estimated by solving an optimization problem, which aims to minimize numerical phase velocity dispersion in homogeneous media for a series of plane waves of different incidence and azimuth angles. This finite difference method leads also to a 27-point stencil involving the eight cells surrounding the central grid (or collocation) point. \\
To deal with heterogeneous media (i.e., variable wavenumber $k$), these weights are generally estimated such that numerical dispersion is jointly minimized for several numbers of grid points per wavelength ($G$). However, the accuracy of the resulting stencil is suboptimal because the same weights are used at each grid point (i.e., row of the matrix) instead of being matched to the local wavelength.
This paper proposes to overcome this accuracy issue by designing a wavelength-adaptive 27-point finite-difference stencil. The straightforward principle first consists of finely tabulating the range of $G$ commonly found in the Earth's crust. Then, the corresponding table of weights is built by minimizing the dispersion for each tabulated $G$, which are processed separately although some regularization can be implemented to force smooth variations of the weights with $G$. Then, each row of the impedance matrix is built by picking in the table the weights corresponding to the local wavelength. A similar idea was proposed for the 2D 9-point stencil by \citet{Xu_2018_AFF}, who concluded from basic 2D simulations that the 9-point adaptive stencil reaches the same accuracy as the 25-point counterpart \citep{Shin_1998_FSD}. However, this study lacks comprehensive validation of the method against structurally complex and contrasted media. \\
The key contribution of this study is to assess the accuracy and robustness of the adaptive 27-point stencil against 3D tectonically complex models involving sharp contrasts for FWI applications. To achieve this goal, we present several large-scale 3D numerical experiments, which show that the accuracy of the adaptive stencil is significantly improved relative to the non-adaptive counterpart without generating any computational overhead. 

This paper is organized as follows. We first briefly review the principles of the 27-point stencil to introduce the different stiffness matrices involved in the stencil and the consistent mass matrix. Then, we present the dispersion analysis and the optimization algorithm that is used to estimate the tabulated weights. At the next step, we present five numerical experiments. The method is first validated against analytical solutions in homogeneous media and linear velocity models in which a large number of wavelengths are propagated. Then, we use the 3D SEG/EAGE overthrust and salt models \citep{Aminzadeh_1997_DSO} as well as the deep crustal GO\_3D\_OBS model \citep{Gorszczyk_2021_GNT} to validate our method in heterogeneous media by comparison with the highly-accurate solutions of the volume-integral convergent Born series (CBS) method \citep{Osnabrugge_2016_CBS}. With the 3D salt benchmark, we also show that the adaptive stencil outperforms a high-order finite-difference time-domain method in the presence of sharp contrasts.


\section{Method}

\subsection{A review of the 27-point finite-difference frequency-domain stencil}
\begin{figure}
\center
\includegraphics[width=0.6\columnwidth,clip=true,trim=0cm 0cm 0cm 0cm]{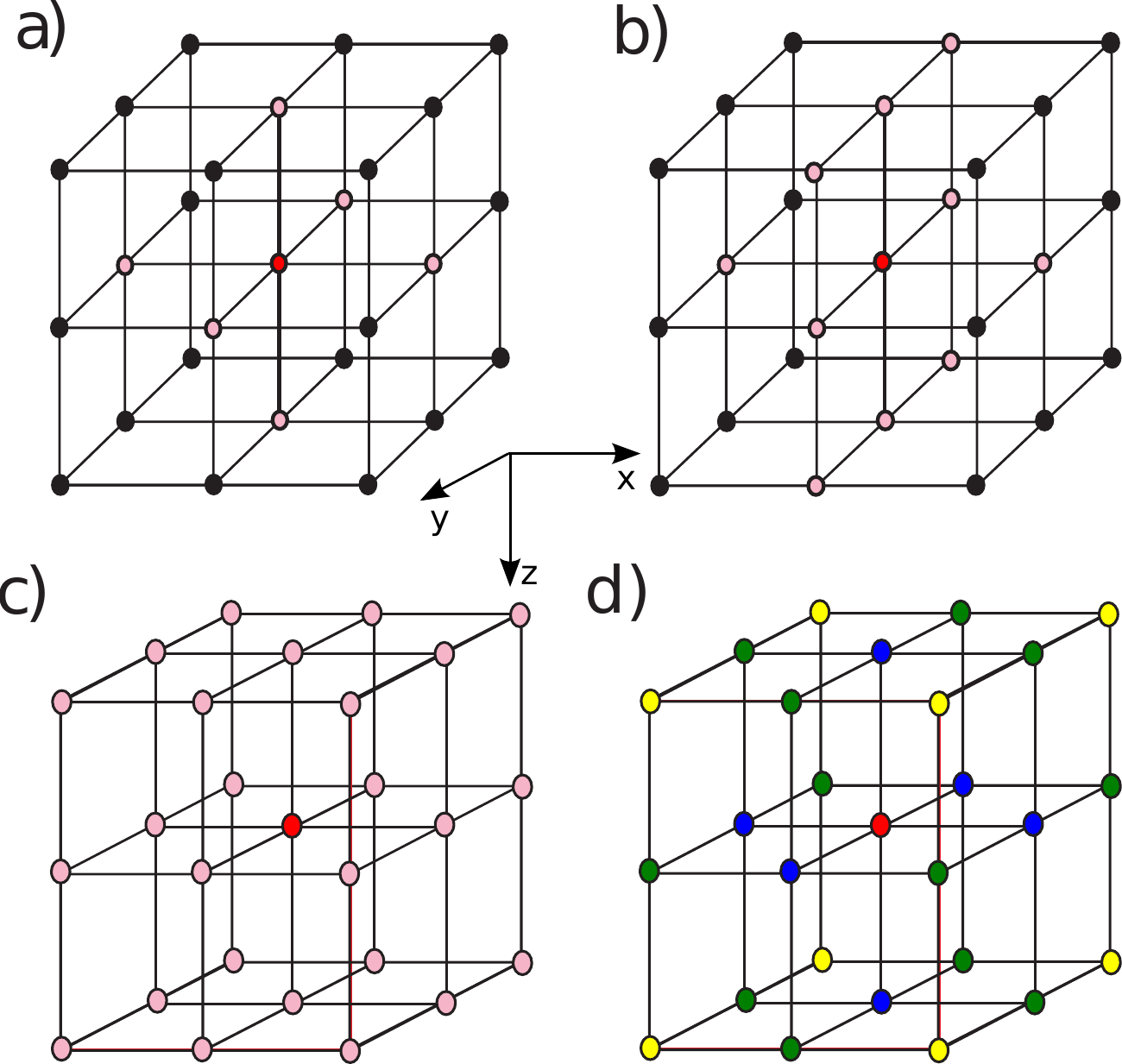}
\caption{Geometry of the 3D 27-point stencil. (a-c) Discretization of stiffness matrix. Circles are pressure grid points. Red circle is the central (collocation) point. Gray circles are points involved in
the stencil in addition to the central point. a) Seven-point stencil on the classic Cartesian coordinate system.
b) Eleven-point stencil obtained by 45$^{o}$ rotation around $x$. The same strategy is applied by rotation around $y$ and $z$. Averaging of the three resulting stencils defines
a 19-point stencil.
c) Stencil obtained from four coordinate systems, each of them being associated with three
main diagonals of a cubic cell. This stencil incorporates 27 coefficients. (d) Categories of grid points are involved in the consistent mass matrix. (Green) Collocation point. (Blue) Points involved in (a). (Red) Points involved in (b). (Yellow) Corner points involved in (c) \citep{Operto_2007_FDFD,Brossier_2010_FNM}.}
\label{fig_coordinate}
\end{figure}

We briefly review the main principles underlying the 27-point finite-difference frequency-domain stencil on a regular Cartesian grid.  
The readers are referred to \citet{Operto_2007_FDFD,Brossier_2010_FNM,Operto_2014_FAT} for more details.\\
According to the parsimonious staggered-grid formulation of \citet{Hustedt_2004_MGS}, we start from the 3D frequency-domain visco-acoustic velocity-stress wave equation
\begin{eqnarray}
&&-i \omega  p(x,y,z,\omega)   =  \kappa(x,y,z) \left( \frac{\partial v_x(x,y,z,\omega)}{\partial \tilde{x}} + \frac{\partial v_y(x,y,z,\omega)}{\partial \tilde{y}}  + \frac{\partial v_z(x,y,z,\omega)}{\partial \tilde{z}} \right),   \label{eq1.1} \\
&&-i \omega v_x(x,y,z,\omega)   =  b(x,y,z)  \frac{\partial p(x,y,z,\omega)}{\partial \tilde{x}} + f_x(x,y,z,\omega),   \label{eq1.2} \\
&&-i \omega  v_y(x,y,z,\omega)  =  b(x,y,z)  \frac{\partial p(x,y,z,\omega)}{\partial \tilde{y}} + f_y(x,y,z,\omega), \label{eq1.3} \\
&&-i \omega v_z(x,y,z,\omega)  =  b(x,y,z)  \frac{\partial p(x,y,z,\omega)}{\partial \tilde{z}}, + f_z(x,y,z,\omega), \label{eq1.4}
\end{eqnarray}
where $\omega$ is the angular frequency, $\kappa(x,y,z)$ is the bulk modulus, $b(x,y,z)$ is the buoyancy (the inverse of density), $p(x,y,z,\omega)$ is the pressure, $v_x(x,y,z,\omega)$, $v_y(x,y,z,\omega)$, $v_z(x,y,z,\omega)$ are the components of the particle velocity vector and $f_x$, $f_y$ and $f_z$ are external forces acting as the source term. We introduce perfectly-matched layer (PML) absorbing conditions in equations~\ref{eq1.1}-\ref{eq1.4} through a change of coordinates in the complex space such that $\partial_{\tilde{x}} = \frac{1}{\xi_x} \, \partial_{x}$, $\partial_{\tilde{y}} = \frac{1}{\xi_y} \, \partial_{y}$, and $\partial_{\tilde{z}} = \frac{1}{\xi_z} \, \partial_{z}$. We have $\xi_x = 1 + i \frac{\gamma_x}{\omega}$, $\xi_y = 1 + i \frac{\gamma_y}{\omega}$ and $\xi_z = 1 + i \frac{\gamma_z}{\omega}$, where the functions $\gamma_x$, $\gamma_y$ and $\gamma_z$ control the damping of the wavefield in the PMLs \citep[][ Their Appendix A]{Operto_2007_FDFD}. \\
Plugging $v_x$, $v_y$ and $v_z$ from equations \ref{eq1.2}-\ref{eq1.4} in equation \ref{eq1.1} gives a generalization of the Helmholtz equation for heterogeneous density:
%
%
\begin{equation}
\frac{\omega^2}{\kappa({\bf{x}})}  p({\bf{x}},\omega) +  \frac{\partial}{\partial x} b({\bf{x}}) \frac{\partial p({\bf{x}},\omega)}{\partial x} 
+ \frac{\partial}{\partial y} b({\bf{x}}) \frac{\partial p({\bf{x}},\omega)}{\partial y} + \frac{\partial }{\partial z} b({\bf{x}}) \frac{\partial p({\bf{x}},\omega)}{\partial z} = s({\bf{x}},\omega),
\label{eq2}
\end{equation}
where ${\bf{x}}=(x,y,z)$ and $s({\bf{x}},\omega) = \nabla \cdot {\bf{f}}$ denotes the pressure source. \\
The above-mentioned elimination procedure applied on the continuous form of the wave equation can be used to discretize equation  \ref{eq2} from the well-documented finite-difference approximation of the first derivative operators involved in equations \ref{eq1.1}-\ref{eq1.4}.
Accordingly, the discretization procedure of equation \ref{eq2} begins with the discretization of the spatial derivatives in equation \ref{eq1.1} with classical second-order accurate staggered-grid finite-difference stencils \citep{Virieux_1984_SWP,Saenger_2000_MPE}. Then, the discretized particle velocities involved in the right-hand side of equation \ref{eq1.1} are developed as a function of the pressure by discretizing equations \ref{eq1.2}-\ref{eq1.4}.
Then, the resulting expressions of the particle velocities are eliminated from the equation \ref{eq1.1} to end up with the discrete form of equation \ref{eq2}. This parsimonious discretization procedure is applied separately on the three different rotated coordinate systems illustrated in Figure~\ref{fig_coordinate}, namely the Cartesian coordinate system leading to the classical seven-point stencil with a cross shape (Figure~\ref{fig_coordinate}a), three rotated coordinate systems obtained by a 45$^\circ$ rotation around each Cartesian axis leading to three 11-point stencils, the combination of them forming a 19-point stencil (Figure~\ref{fig_coordinate}b), and four rotated coordinate systems defined by three diagonals of the cubic cell leading to four 27-point stencils (Figure~\ref{fig_coordinate}c).
Note also that this elimination procedure can be applied in a simpler setting by considering the approach of \citet{Min_2000_IFD} extended to the 3D acoustic and elastic equations by \citet{Chen_2012_DMF} and \citet{Gosselin_2014_FDF}, respectively. In this approach, the classical second-order stencil is used to discretize the spatial derivatives in the three Cartesian directions at the collocation point but also at the neighbouring points such that the support of the mixed stencil spans the 27 points surrounding the collocation point. \\
After discretization, the generalized Helmholtz, equation \ref{eq2}, can be written in matrix form as
\begin{equation}
\left[\bold{M} + \bold{S} \right] \bold{p} = \bold{s},
\end{equation}
where $\bold{M}$ and  $\bold{S}$ are the mass and stiffness matrices, respectively. \\
The stiffness matrix $\bold{S}$ is formulated as a weighted sum of the stiffness matrices formulated on each coordinate system
\begin{equation}
\bold{S}=w_{s_1} \bold{S}_1 + \frac{w_{s_2}}{3} \sum_{i=1}^3 \bold{S}_{2,i} + \frac{w_{s_3}}{4} \sum_{i=1}^4 \bold{S}_{3,i},
\end{equation}
where the weights satisfy
\begin{equation}
\sum_{i=1}^3 w_{s_i} = 1,
\label{eqsws}
\end{equation}
and $\bold{S}_1$, $\bold{S}_{2,\bullet}$ and $\bold{S}_{3,\bullet}$ denote the stiffness matrices associated with the three classes of coordinate system illustrated in Figure~\ref{fig_coordinate}. \\

The consistent mass matrix $\bold{M}$ is built by distributing the term $\frac{\omega^2}{\kappa({\bf{x}})}$ over the 27 points as
\begin{equation}
\frac{\omega^2}{\kappa} p \Longrightarrow \omega^2 \left( w_{m_0} p_0
+ \frac{w_{m_1}}{6} \sum_{i=1}^6 \left[\frac{p}{\kappa}\right]_{1,i}+ \frac{w_{m_2}}{12} \sum_{i=1}^{12} \left[\frac{p}{\kappa}\right]_{2,i} +\frac{w_{m_3}}{8} \sum_{i=1}^8 \left[\frac{p}{\kappa}\right]_{3,i} \right) ,
\label{eq8}
\end{equation}
where the weights satisfy
\begin{equation}
\sum_{i=0}^3 w_{m_i}= 1,
\label{eqswm}
\end{equation}
and the subscript $0$ denotes the collocation point (Figure~\ref{fig_coordinate}d, green point), the subscripts $1,\bullet$ denote the 6 neighbour grid points involved in the classical 7-point stencil (Figure~\ref{fig_coordinate}d, blue points), the subscripts $2,\bullet$ denote the 12 additional grid points involved in the three stencils of type 2 (Figure~\ref{fig_coordinate}d, red points) and the subscripts $3,\bullet$ denote the eight additional grid points at the corners involved in the four stencils of type 3 (Figure~\ref{fig_coordinate}d, yellow points).
Alternatively, one may use for the consistent mass
\begin{equation}
\frac{\omega^2}{\kappa} p \Longrightarrow \frac{\omega^2}{\kappa_0} \left( w_{m_0} p_0 + \frac{w_{m_1}}{6} \sum_{i=1}^6 p_{1,i}+ \frac{w_{m_2}}{12} \sum_{i=1}^6 p_{2,i} +\frac{w_{m_3}}{8} \sum_{i=1}^8 p_{3,i} \right) ,
\label{eq8}
\end{equation}
where the bulk modulus at the collocation point is used for all the 27 coefficients. This later formulation preserves the bilinearity property of the wave equation \citep{Aghamiry_2019_AMW,Aghamiry_2019_IWR}.

\subsection{Numerical dispersion analysis for adaptive mass and stiffness weights}
The dispersion analysis in homogeneous media of the 27-point stencil is presented in \citet{Operto_2007_3D}.
Plugging a plane wave in the discretized wave equation gives the following expression of the numerical phase velocity normalized by the wavespeed $c$.
\begin{equation} \label{vtilde}
\tilde{v}=\frac{G}{\sqrt{2J}\pi}\sqrt{w_{s_1}(3-C)+\frac{w_{s_2}}{3}(6-C-B)+\frac{2w_{s_3}}{4}(3-3A+B-C)},
\end{equation} 
where $G$ is the number of points per wavelength $\lambda$, $J=(w_{m_0}+2w_{m_1}C+4w_{m_2} B+8w_{m_3} A)$, and the functions $A$, $B$  and $C$ are given by 
\begin{equation}
\begin{cases}
A(G,\theta,\phi)=\cos(\frac{2\pi \cos(\phi)\cos(\theta)}{G})\cos(\frac{2\pi\cos(\phi)\sin(\theta)}{G})\cos(\frac{2\pi \sin(\phi)}{G}), \\
B(G,\theta,\phi)=\cos(\frac{2\pi \cos(\phi)\cos(\theta)}{G})\cos(\frac{2\pi\cos(\phi)\sin(\theta)}{G})+\cos(\frac{2\pi \cos(\phi)\cos(\theta)}{G})\cos(\frac{2\pi\sin(\phi)}{G}) \\
\hspace{8cm}+\cos(\frac{2\pi \cos(\phi)\sin(\theta)}{G})\cos(\frac{2\pi \sin(\phi)}{G}),\\
C(G,\theta,\phi)=\cos(\frac{2\pi \cos(\phi)\cos(\theta)}{G})+\cos(\frac{2\pi\cos(\phi)\sin(\theta)}{G})+\cos(\frac{2\pi \sin(\phi)}{G}),
\end{cases}
\end{equation}
where $\phi \in [0~ \pi/4]$ and $\theta \in [0 ~\pi/2]$ are the incidence and azimuth angles of the plane wave.

\begin{figure}
\center
\includegraphics[width=1\columnwidth,clip=true,trim=0.5cm 18.5cm 0.8cm 3.5cm]{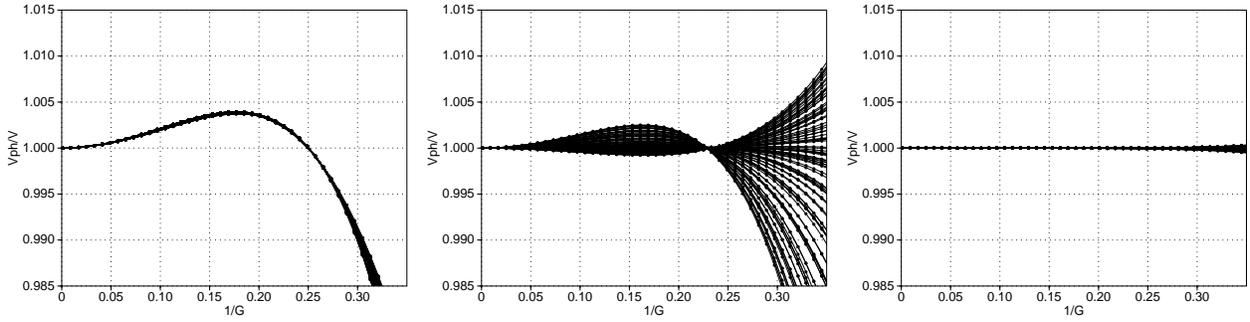}
\caption{Phase velocity dispersion curves showing the ratio between the numerical phase velocity and the wavespeed versus $1/G$. (a) Weights of the 27-point mixed-grid stencil are estimated to minimize dispersion for $G$=4, (b) Same as (a) for $G$=4,6,8,10. (c) Adaptive weights are estimated separately for a fine table of $1/G$ ranging between 0 and 0.4 with a step of 0.001. The dip and azimuth angles of the plane waves range between 0 and 45$^o$ with a step of 10$^o$.  }
\label{fig_dispersion}
\end{figure}

\begin{figure}
\center
\includegraphics[width=1.0\columnwidth,clip=true,trim=1.8cm 8cm 0cm 5cm]{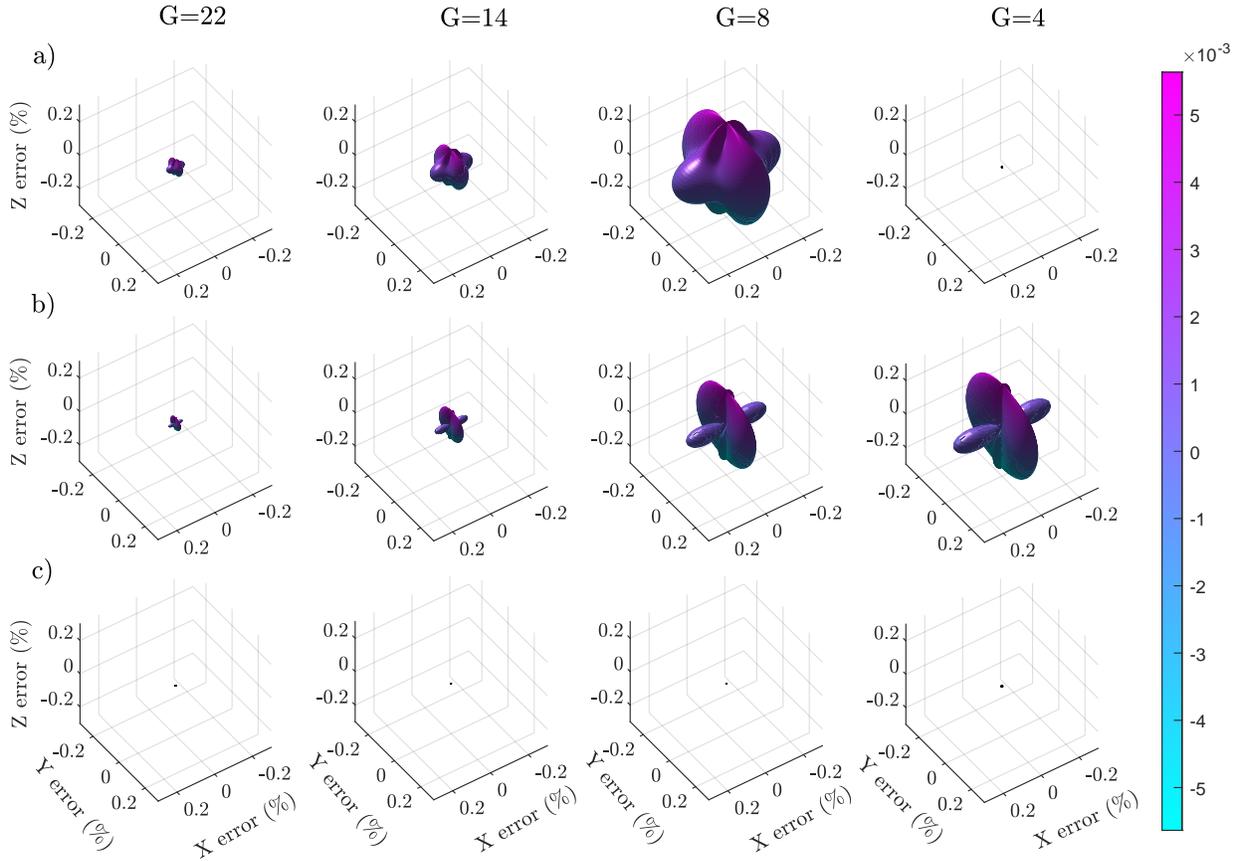}
\caption{Three-dimensional view of the numerical phase velocity error (in percent). Numerical dispersion and anisotropy are illustrated. (a) The numerical phase velocity errors are computed with the weights minimizing dispersion for $G$=4. From left to right, error for $G$=4,6,8, and 10. (b) Same as (a) when the weights are computed to jointly minimize dispersion for $G$=4, 6, 8, and 10. (c) Same as (a) for the adaptive weights.}
\label{fig_dispersion1}
\end{figure}

Considering the relationships in equations \ref{eqsws} and \ref{eqswm}, there exist only five independent weights that are to be determined for minimizing the numerical dispersion. These weights satisfy the following linear equation (obtained by requiring $\tilde{v}=1$, eliminating $w_{s_3}$ and $w_{m_4}$ and rearranging the terms in equation \ref{vtilde}):
\begin{equation} \label{linear_basic}
\sum_{l=1}^5 H_l(G,\theta,\phi) w_l  = g(G,\theta,\phi),
\end{equation}
where $\bold{w}=[w_{s_1}, w_{s_2},w_{m_0}, w_{m_1}, w_{m_2}]^T$,
$g(G,\theta,\phi)= \frac12((\frac{4\pi^2}{ G^2}+ 3)A - B + C-3)$ and
\begin{equation}
\begin{cases}
H_1(G,\theta,\phi)= \frac12(3 + 3A - B - C),\\
H_2(G,\theta,\phi)= \frac12(3A - \frac{5}{3}B + \frac{1}{3}C),\\
H_3(G,\theta,\phi)= \frac{2\pi^2}{ G^2}(A - 1),\\
H_4(G,\theta,\phi)= \frac{2\pi^2}{ G^2}(6A - 2C),\\
H_5(G,\theta,\phi)= \frac{2\pi^2}{ G^2}(12A - 4 B).
\end{cases}
\end{equation}
In order to solve equation \ref{linear_basic},  we define the following misfit function
\begin{equation} \label{eq_cost}
\sum_{i=1}^{N_G} \sum_{j=1}^{N_\theta} \sum_{k=1}^{N_\phi} 
\left(\sum_{l=1}^5 H_l^{(ijk)} w_l - g^{(ijk)} \right)^2,
\end{equation} 
where $\bullet^{(ijk)}\equiv \bullet(G_i,\theta_j,\phi_k)$.
This function can be minimized for one value of $G$ ($N_G$=1) if an accurate monochromatic simulation should be performed in a homogeneous medium. The phase velocity dispersion curves as a function of $1/G$  are shown in Figures~\ref{fig_dispersion}a, when the weights have been optimized for $G$ = 4. We also plot phase velocity dispersion surfaces in spherical coordinates \citep[][ Their Figure 3]{Brossier_2010_FNM} for different values of $G$ to better assess the numerical anisotropy of the stencil (Figure~\ref{fig_dispersion1}a). A high accuracy is achieved for $G$=4. However, significant errors are shown for higher values of $G$ (lower frequencies), with the maximum error reached at G $\approx$ 5.5 suggesting that this setting is suboptimal for simulation in heterogeneous media. 

To improve accuracy in heterogeneous media, \citet{Operto_2007_3D} and \citet{Brossier_2010_FNM} minimize the equation \ref{eq_cost} for a range of $G$ representative of the wavelengths propagated in the subsurface medium during a monochromatic simulation. More precisely, they minimize the misfit function for $N_G$ = 4 with $G$=4, 6, 8 and 10. The corresponding dispersion curves in Figure~\ref{fig_dispersion}a show that the phase velocity errors have been mitigated in an average sense, although the minimum error is higher compared to the previous case (Figure~\ref{fig_dispersion1}b). 

\begin{figure}
\center
\includegraphics[width=0.75\columnwidth,clip=true,trim=0cm 2.5cm 0cm 5cm]{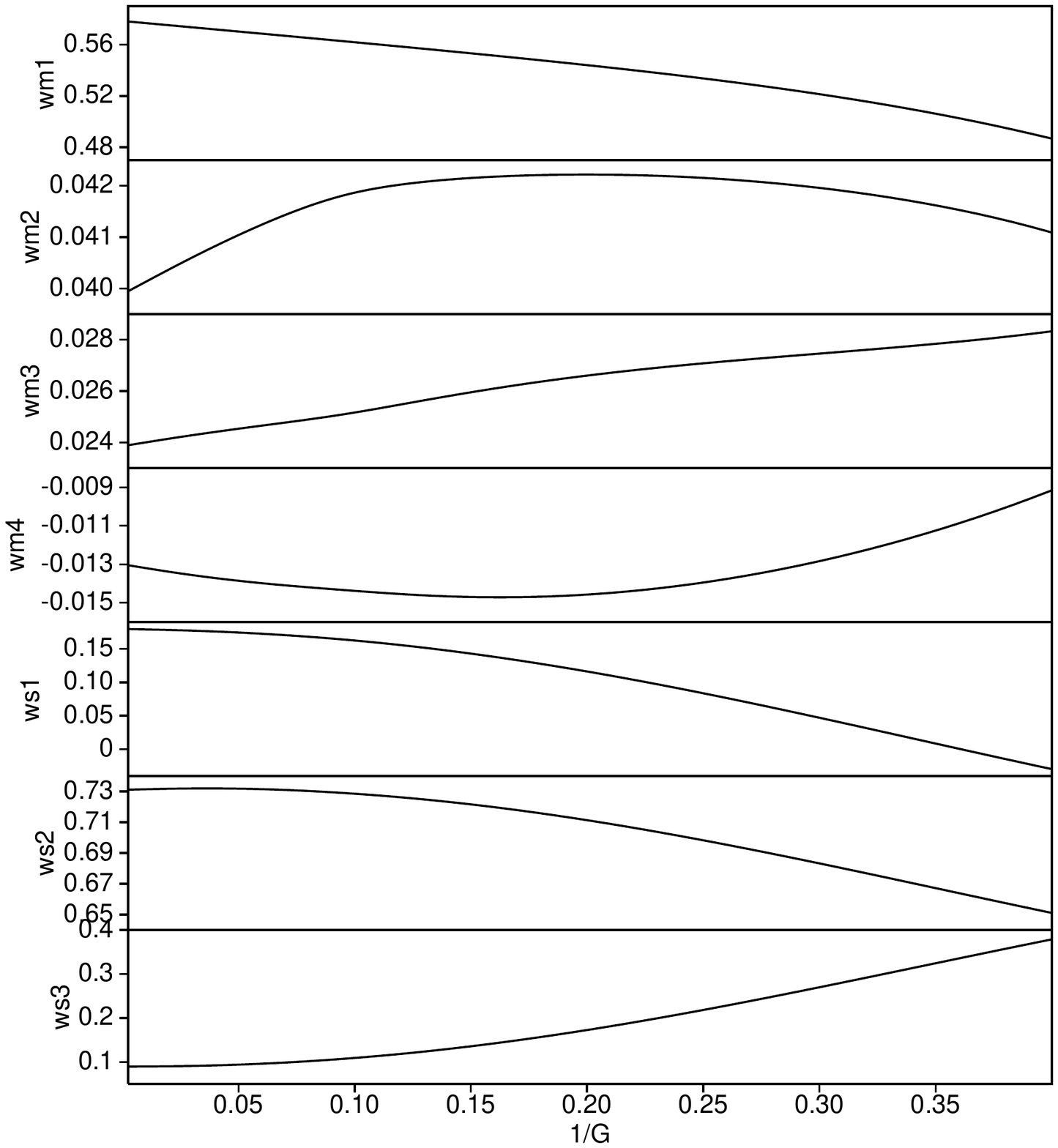}
\caption{Tabulated values of the stiffness-matrix weights ($w_{s_1}$,$w_{s_2}$,$w_{s_3}$) and the mass-matrix weights ($w_{m_1}$,$w_{m_2}$,$w_{m_3}$,$w_{m_4}$) as a function of 1/G. }
\label{fig_coeff}
\end{figure}

In both approaches, the 27-point stencil is non-adaptive since the weights don't depend on the local wavelength.
To implement wavelength adaptivity, first we build the weights as functions of $G$. The set of weight functions
\begin{equation}
\{w_{s_1}(G),w_{s_2}(G),w_{m_0}(G),w_{m_1}(G),w_{m_2}(G)\},
\end{equation}
can be built by using either a parametric or non-parametric regression via equation \ref{eq_cost}.
In a non-parametric setting, equation \ref{eq_cost} reads
\begin{equation} \label{eq_cost_fun}
\sum_{i=1}^{N_G} \sum_{j=1}^{N_\theta} \sum_{k=1}^{N_\phi} 
\left(\sum_{l=1}^5 H_l^{(ijk)} w_l^{(i)} - g^{(ijk)} \right)^2,
\end{equation}
where $\bold{w}^{(i)}=[w_{s_1}^{(i)}, w_{s_2}^{(i)},w_{m_0}^{(i)}, w_{m_1}^{(i)}, w_{m_2}^{(i)}]^T$. \\
We now describe in more detail the optimization algorithm that we use to estimate the weights.
Equation \ref{eq_cost_fun} is separable and may be solved for each $i$ separately as
\begin{equation} \label{OPFD}
\underset{\substack{\bold{w}^{(i)}}}{\text{minimize}} ~\|\bold{H}^{(i)}\bold{w}^{(i)} - \bold{g}^{(i)}\|_2^2,
\end{equation} 
where 
\hspace{1cm}
\begin{equation}
\bold{H}^{(i)}=
\begin{bmatrix}
H_1^{(i11)} & H_2^{(i11)} & H_3^{(i11)} & H_4^{(i11)} & H_5^{(i11)}\\
H_1^{(i21)} & H_2^{(i21)} & H_3^{(i21)} & H_4^{(i21)} & H_5^{(i21)}\\
\vdots & \vdots & \vdots & \vdots & \vdots\\
H_1^{(iN_{\theta}1)} & H_2^{(iN_{\theta}1)} & H_3^{(iN_{\theta}1)} & H_4^{(iN_{\theta}1)} & H_5^{(iN_{\theta}1)}\\
H_1^{(iN_{\theta}2)} & H_2^{(iN_{\theta}2)} & H_3^{(iN_{\theta}2)} & H_4^{(iN_{\theta}2)} & H_5^{(iN_{\theta}2)}\\
\vdots & \vdots & \vdots & \vdots & \vdots\\
H_1^{(iN_{\theta}N_{\phi})} & H_2^{(iN_{\theta}N_{\phi})} & H_3^{(iN_{\theta}N_{\phi})} & H_4^{(iN_{\theta}N_{\phi})} & H_5^{(iN_{\theta}N_{\phi})}\\
\end{bmatrix}, ~~~~
\bold{g}^{(i)}
=
\begin{bmatrix}
g^{(i11)}\\
g^{(i21)}\\
\vdots \\
g^{(iN_{\theta}1)}\\
g^{(iN_{\theta}2)}\\
\vdots \\
g^{(iN_{\theta}N_{\phi})}\\
\end{bmatrix}.
\end{equation}
However, solving these overdetermined systems independently without any prior information about the weights might cause significant oscillatory variations of each weight as a function of $G$, which may be undesired for accurate wavefield estimation.
Therefore, we estimate the $N_G$ series of weights simultaneously by gathering the $\bold{H}^{(i)}$ matrices in a block diagonal matrix $\bold{H}$ of dimension $(N_{\theta} \times N_{\phi} \times N_G) \times (5 \times N_G)$ and by augmenting the misfit function with a Tikhonov smoothing regularizer to enforce smooth variations of each weight as a function of $G$. 
Accordingly, the regularized optimization problem reads
\begin{eqnarray}
\underset{\substack{\bold{w}}}{\text{minimize}} ~\|\bold{H}\bold{w}-\bold{g}\|_2^2+\lambda\|\bold{L}\bold{w}\|_2^2,
\end{eqnarray}
where
\begin{equation}
\bold{H}=\begin{bmatrix}
\bold{H}^{(1)} &  &\\  
&  \ddots & \\
 &   & \bold{H}^{(N_G)}
\end{bmatrix},~~~
\bold{g}=\begin{bmatrix}
\bold{g}^{(1)} \\  
\vdots\\
\bold{g}^{(N_G)}
\end{bmatrix},~~~
\bold{w}=\begin{bmatrix}
\bold{w}^{(1)} \\  
\vdots\\
\bold{w}^{(N_G)}
\end{bmatrix},
\end{equation}
and  $\bold{L}$ is a smoothing operator. Note that the values of not shown coefficients in matrix $\bold{H}$ are zero. One may build $\bold{L}$ as the product of two matrices: $\bold{L=DP}$ where $\bold{D}$ is a difference operator with appropriate boundary conditions and $\bold{P}$ is a permutation matrix which reorders the coefficients of $\bold{w}$ such that the first $N_G$ entries are $[w_{s_1}^{(1)}, ...,w_{s_1}^{(N_G)}]$, the second $N_G$ entries are $[w_{s_2}^{(1)}, ...,w_{s_2}^{(N_G)}]$ and so on.  The  smooth variation of the estimated adaptive weights as a function of $G$ are shown in Figure~\ref{fig_coeff} while the dispersion curves and surfaces of the adaptive stencil are shown in Figures~\ref{fig_dispersion}c and \ref{fig_dispersion1}c. As expected, the numerical dispersion can be theoretically reduced very significantly for a wide range of $G$ by treating each $G$ separately during the dispersion analysis. \\
The goal of the next section is to check the robustness of this approach with large-scale numerical experiment involving complex structures and sharp material discontinuities.

\section{Numerical examples}

\subsection{Experimental setup}
\begin{figure}
\center
\includegraphics[width=0.7\columnwidth,clip=true,trim=0cm 0cm 0cm 0cm]{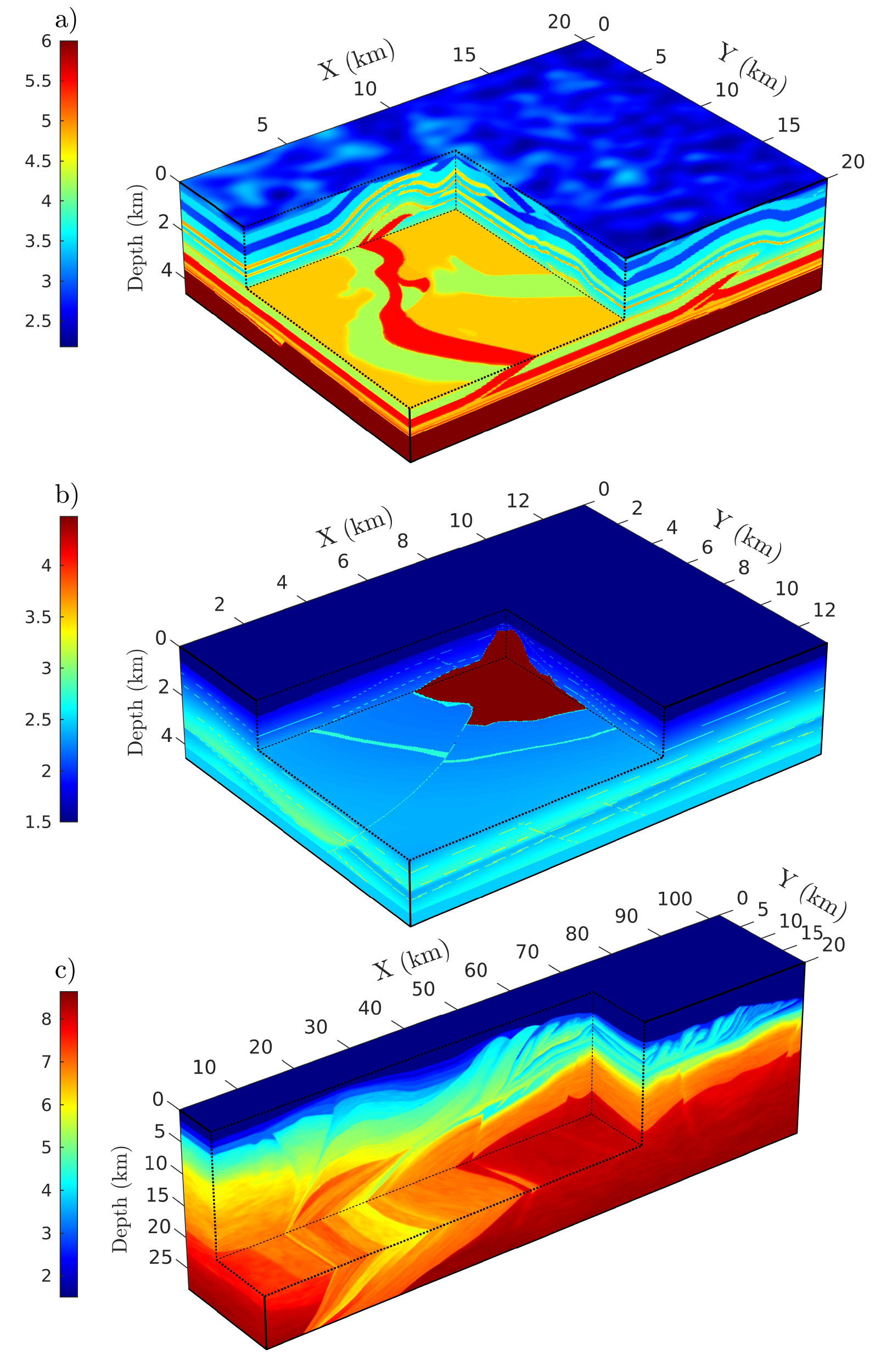}
\caption{Perspective view of the (a) overthrust, (b) salt, and (c) GO\_3D\_OBS models.}
\label{fig_over_salt}
\end{figure}

We validate our method against five benchmarks (Table~\ref{tab_benchs}). We limit ourselves to isotropic acoustic media with a constant density equal to 1. However, the conclusions of these tests should apply to visco acoustic media with heterogeneous density. It should also apply to acoustic media with transverse isotropic (TI) effects and viscoelastic media. In the first case, the normalized phase velocity for the quasi-P mode is a function of the vertical wavespeed and the Thomsen's parameters $\delta$ and $\epsilon$ \citep{Operto_2009_FDF}. An approximation can also be used to estimate the optimal weights from the local elliptic phase velocity \citep[][ Their equation 15]{Operto_2014_FAT} In the elastic case, the normalized P and S phase velocity can be formulated as a function of the Poisson ratio and the shear wavenumber and hence the optimal weights can be formulated as functions of these quantities \citep{Stekl_1998_AVM,Gosselin_2014_FDF}.  The first two benchmarks involve an infinite homogeneous medium and an infinite laterally-homogeneous medium where the velocity linearly increases in one direction. For both media, the accuracy of the finite-difference frequency-domain (FDFD) solutions is assessed against analytical solutions. The reader is referred to \citet{Kuvshinov_2006_EST} for the analytical solution of the Helmholtz equation in a velocity gradient model. The last three benchmarks involve complex velocity models corresponding to the 3D SEG/EAGE overthrust model (Figure~\ref{fig_over_salt}a), the 3D SEG/EAGE salt model (Figure~\ref{fig_over_salt}b) \citep{Aminzadeh_1997_DSO} and a target of the 3D GO\_3D\_OBS model (Figure~\ref{fig_over_salt}c) \citep{Gorszczyk_2021_GNT}. The two SEG/EAGE models are representative of the exploration geophysics scale. At this scale, $G$ typically ranges between 4 and 12 during a monochromatic simulation.  The 3D GO\_3D\_OBS model covers a continental margin at a regional scale. In this case, $G$ typically ranges between 4 and 25 due to the wider range of wavespeeds found in the crust and upper mantle. For these last three models, the FDFD solution is assessed against a reference solution computed with the highly-accurate Convergent Born Series (CBS) method \citep{Osnabrugge_2016_CBS}. Compared to finite-difference or finite-element methods that rely on discretization of differential operators, the CBS method belongs to another class of methods referred to as volume integral methods based upon the Green's function theorem. More specifically, the solution of the Helmholtz equation in heterogeneous media is inferred from Fourier-domain analytic expression of the Green's functions in homogeneous media by solving the scattered-field wave equation with material contrasts of arbitrary scattering strength starting from the zero-order term of the Born series and iteratively recovering the higher-order scattering terms.
Therefore, we expect to achieve an accuracy of the CBS solutions at the level of machine precision while \citet{Osnabrugge_2016_CBS} conclude that the CBS method is nine orders more accurate than the pseudospectral time-domain method. For all the extracted wavefields using CBS, the stopping criterion of iteration is set according to a backward error of 1e-12. In addition to visual comparison, we assess the accuracy of the adaptive and non-adaptive FDFD stencils by computing the $\ell{1}$ norm of the difference between the CBS and FDFD wavefields (Table~\ref{tab_accuracy}) according to the formula
\begin{equation}
err=\frac{| \bold{W}  \mathcal{R}(\bold{u}_{cbs}-\bold{u}_{fdfd}) |}{ | \bold{W}  \mathcal{R}(\bold{u}_{cbs})|}+\frac{|  \bold{W} \mathcal{I}(\bold{u}_{cbs}- \bold{u}_{fdfd}) |}{  | \bold{W} \mathcal{I}(\bold{u}_{cbs}) |},
\label{eqerror}
\end{equation}
where  $\bold{u}_{cbs}$ and $\bold{u}_{fdfd}$ denotes the CBS and FDFD wavefields, respectively, $\mathcal{R}$ and $\mathcal{I}$ denote the real and imaginary parts of a complex number, respectively, and $\bold{W}$ is a diagonal matrix applying a linear gain with distance from the source to the wavefield for amplitude balancing.

For each benchmark, we perform the FDFD simulation with non-adaptive and adaptive weights. In the non-adaptive case, we use two sets of weights that have been obtained by minimizing dispersion for $G=4$ $(N_G=1)$ and $G=4,6,8,10$ $(N_G=4)$. In the following, the wavefields computed with such weights are referred to as $G4$ and $Gm$ wavefields, respectively. We test two implementations of the adaptive stencil. The first implementation builds each row of the impedance matrix with the weights corresponding to the local wavelength at the collocation point. The second one builds each row of the impedance matrix with average weights, the averaging being performed over the 27 points of the stencil. In the following, the wavefields computed with these two adaptive approaches are referred to as $GA$ and $GAm$ wavefields, respectively.
The source is a point source located on a grid point where the temporal source signature is a delta function, unless otherwise mentioned: $s(\bold{x},\omega)= \delta(\bold{x}-\bold{x}_s)$, where $\delta$ denotes the delta function and $\bold{x}_s$ denotes the source coordinates.  We solve the discretized Helmholtz equation in single precision with the full-rank version of the massively-parallel sparse direct solver MUMPS \citep{MUMPS_2021_MMP} based upon the multifrontal method \citep{Duff_1983_MSI,Amestoy_2001_FAM,Amestoy_2003_APS,Duff_2017_DMS}. We implement PMLs along each face of the grid. We perform the simulation on the Occigen supercomputer of CINES (\url{https://www.cines.fr}). The computer nodes contain two  Haswell E5-2690V3@2.6 GHz processors with 128 Giga bytes of shared memory and 12 cores per processor. The high-speed network is Infiniband FDR 56 Gbit/s. For the simulation, we assign one Message Passing Interface (MPI) per node and use multithreading with 24 threads per MPI process to minimize memory overheads during Lower-Upper (LU) decomposition while optimizing computational time.

\subsection{Homogeneous velocity model}
\begin{figure}
\center
\includegraphics[width=1\columnwidth]{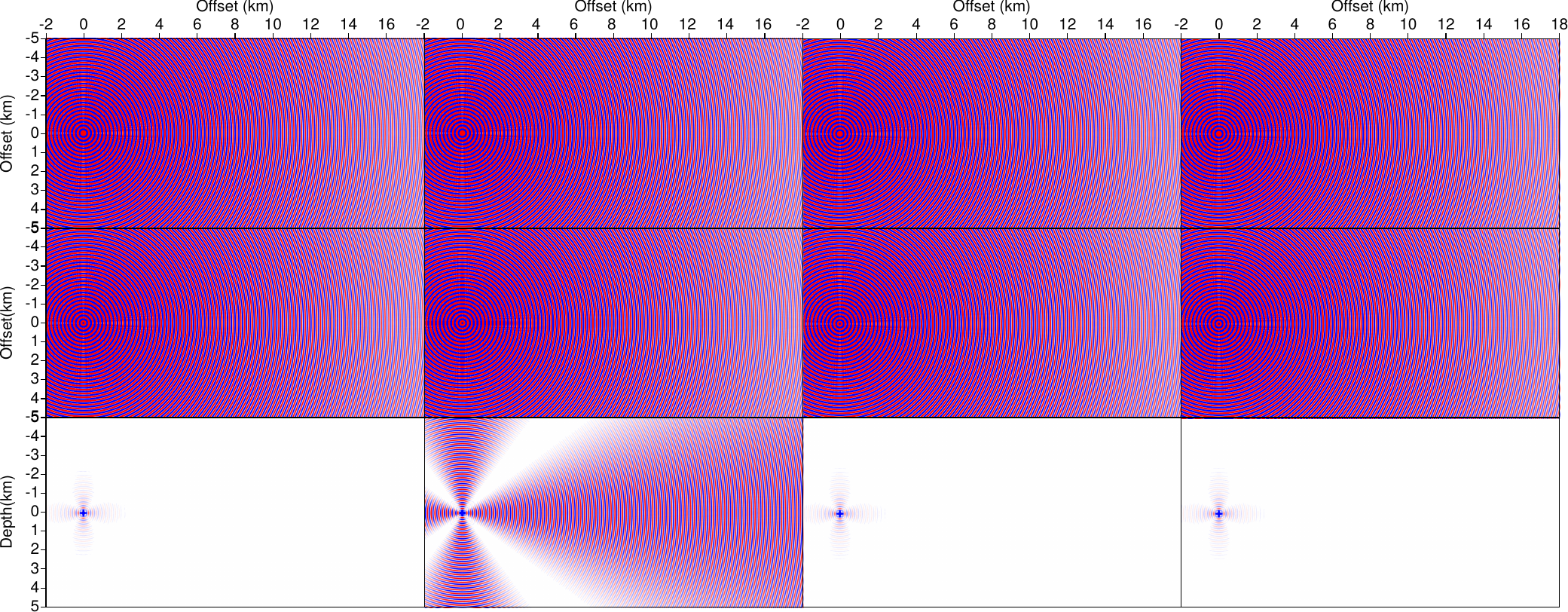}
\caption{Homogeneous model. Top panels show the real part of the analytical solution. Middle panels show, from left to right, the real part of the $G4$, $Gm$, $GA$, $GAm$ wavefields. The bottom panels show the differences between the analytical and FDFD solutions. The vertical and horizontal axis correspond to the $x$ and $y$ dimensions, respectively.}
\label{fig_homogeneous_wavefield}
\end{figure}

\begin{figure}
\center
\includegraphics[width=0.6\columnwidth]{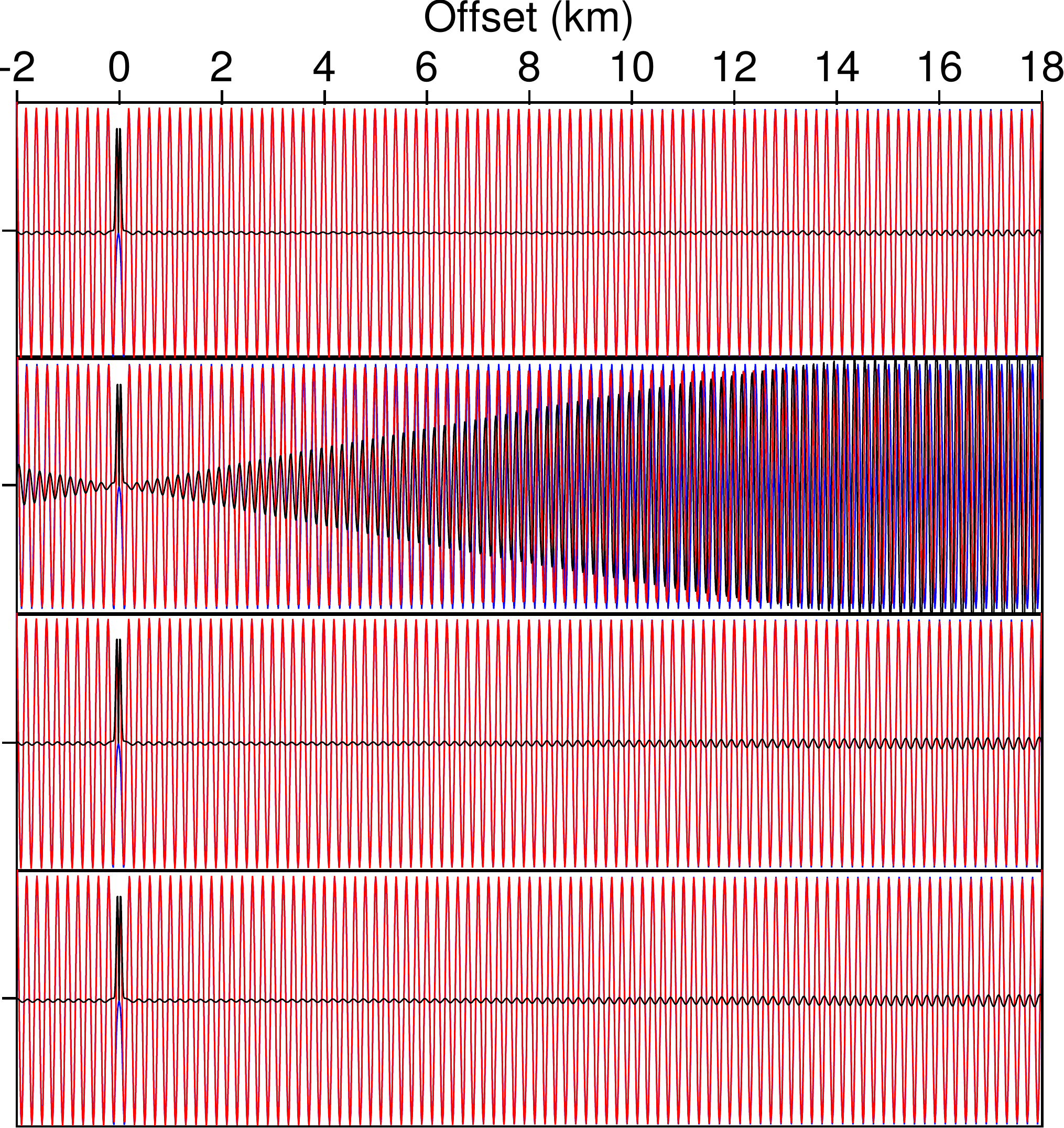}
\caption{Homogeneous model. Direct comparison between analytical (blue) and FDFD (red) solutions (real part) along an horizontal profile across the source oriented in the $y$ direction. The black line is the difference. Amplitudes are corrected for geometrical spreading with a linear gain with offset. From top to bottom, real part of the $G4$, $Gm$, $GA$, $GAm$ wavefields.}
\label{fig_homogeneous_log}
\end{figure}

We first consider a homogeneous medium of size 10~km $\times$ 20~km $\times$ 10~km with a wavespeed of 1500~m/s (Table~\ref{tab_benchs}). The medium is discretized with a grid interval of 50~m, and the frequency is 7.5~Hz, leading to $G$=4. The number of grid points in the PML is as high as 16 to minimize the footprint of parasite reflections in the accuracy assessment of the stencil. The source is positioned at (5~km, 2~km, 5~km). We perform the simulation with 40 nodes of the Occigen computer. The elapsed time and the memory required to perform the LU factorization is 837~s and around 2~Tera bytes. The analytical and FDFD wavefields in the (x,y) plane across the source, as well as the difference between the two solutions, are shown in Figure~\ref{fig_homogeneous_wavefield}. The direct comparison between the two wavefields and the error along the $y$-profile across the source are shown in Figure~\ref{fig_homogeneous_log}. As expected, the $G4$ wavefield is highly accurate since we discretize the homogeneous medium with $G$=4 to perform the simulation. Consistently with the dispersion curves, the accuracy of the $GA$ and $GAm$ wavefields are similar to that of the $G4$ counterpart, while the $Gm$ wavefield is significantly less accurate since the $Gm$ weights result from a compromise to jointly minimize the dispersion for $G$=4,6,8 and 10. 

\subsection{Linear velocity model}
\begin{figure}
\center
\includegraphics[width=1\columnwidth]{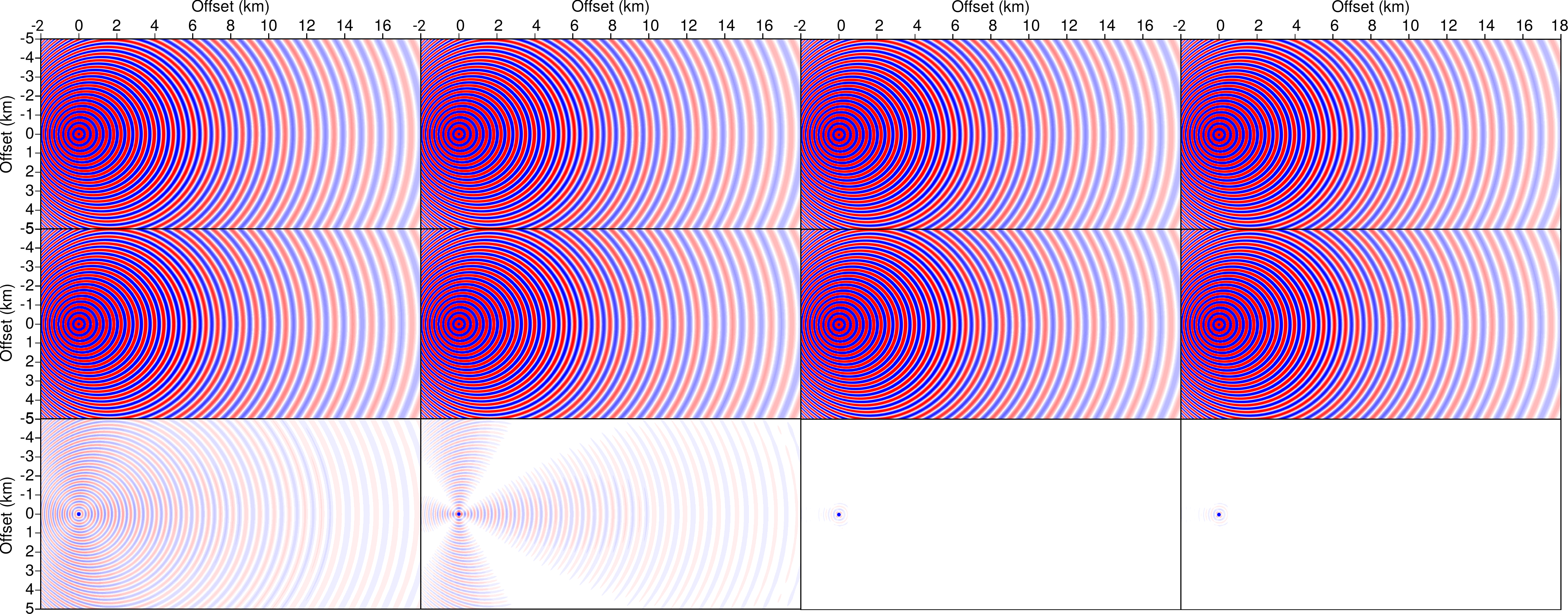}
\caption{Linear velocity model. Top row: Real part of the analytical solution. Middle row from left to right: Real part of the wavefield computed with $G4$, $Gm$, $GA$, and $GAm$ weights. Bottom row: Differences between analytical and FDFD wavefields. The vertical and horizontal axis represents the $x$ and $y$ directions, respectively.}
\label{fig_wavefield_gradient}
\end{figure}

\begin{figure}
\center
\includegraphics[width=0.6\columnwidth]{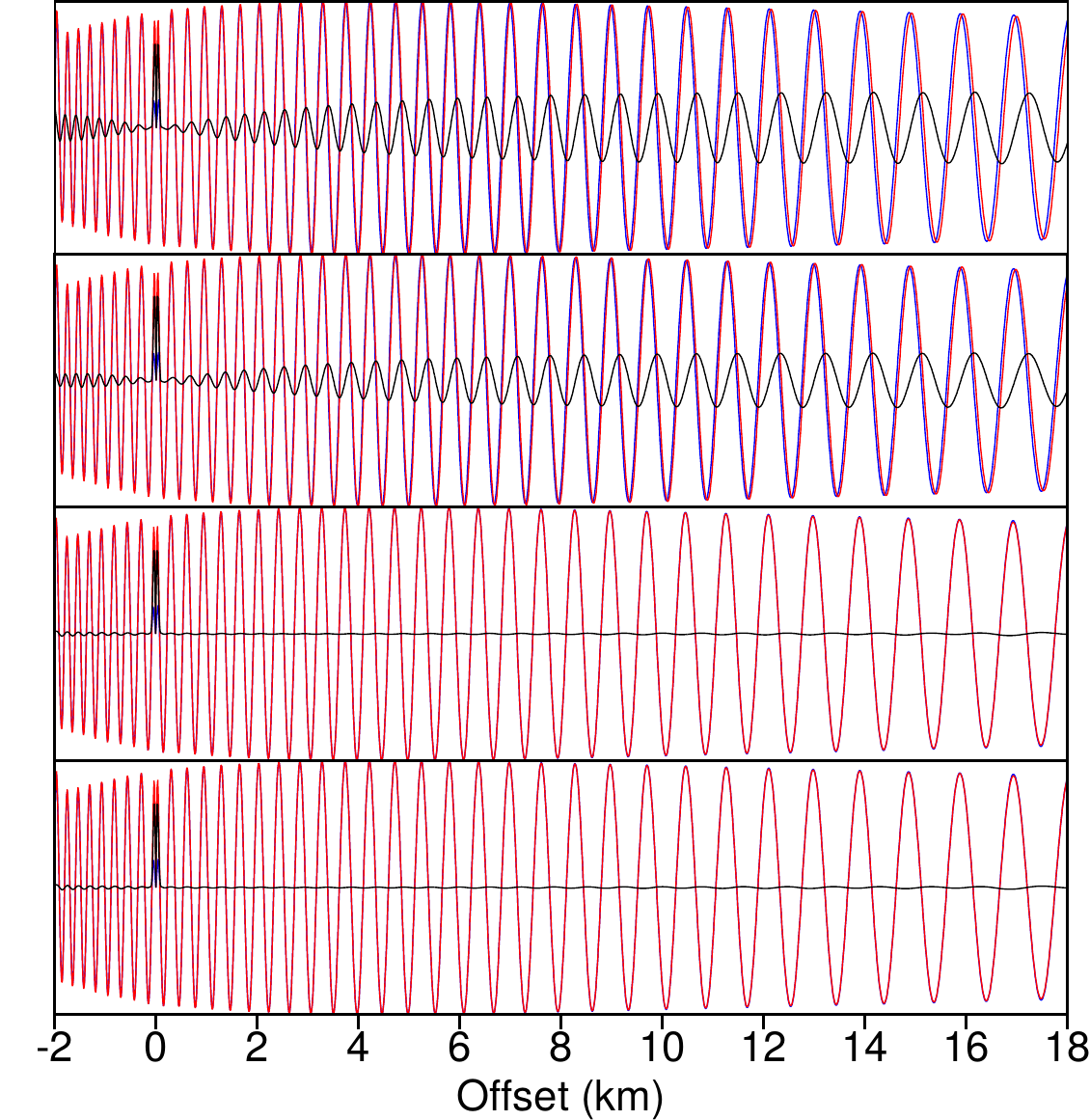}
\caption{Linear velocity model. Direct comparison between analytical (blue) and FDFD (red) wavefield solutions (real part) along an horizontal profile across the source oriented in the $y-$direction. The black line is the difference. Amplitudes are corrected for geometrical spreading with a linear gain with offset. From top to bottom, $G4$, $Gm$, $GA$, $GAm$ wavefields.}
\label{fig_log_gradient}
\end{figure}

The second benchmark involves a laterally homogeneous medium in which the velocity linearly increases in the $y$ direction. The size of the grid and the frequency are the same as that used for the homogeneous medium. The wavespeed increases from 1500~m/s at y=0~km to 8500~m/s at y=20~km hence covering the wavespeeds encountered in the earth's crust and upper mantle (the velocity gradient is $0.35$ s$^{-1}$). Accordingly, $G$ increases from 4 to $\sim$23 (Table~\ref{tab_benchs}). The results are shown in Figures~\ref{fig_wavefield_gradient} and \ref{fig_log_gradient} with the same showing as for the homogeneous case. As expected, the $G4$ wavefield accumulates phase error as $G$ increases in the $y$ direction. The $Gm$ wavefield shows a better accuracy as expected, while the $GA$ and $GAm$ wavefields show a higher accuracy without visible phase error and small amplitude errors, hence providing a first validation of the adaptive stencil in smooth media.

\subsection{3D SEG/EAGE Overthrust model}
\begin{figure}
\center
\includegraphics[width=1\columnwidth]{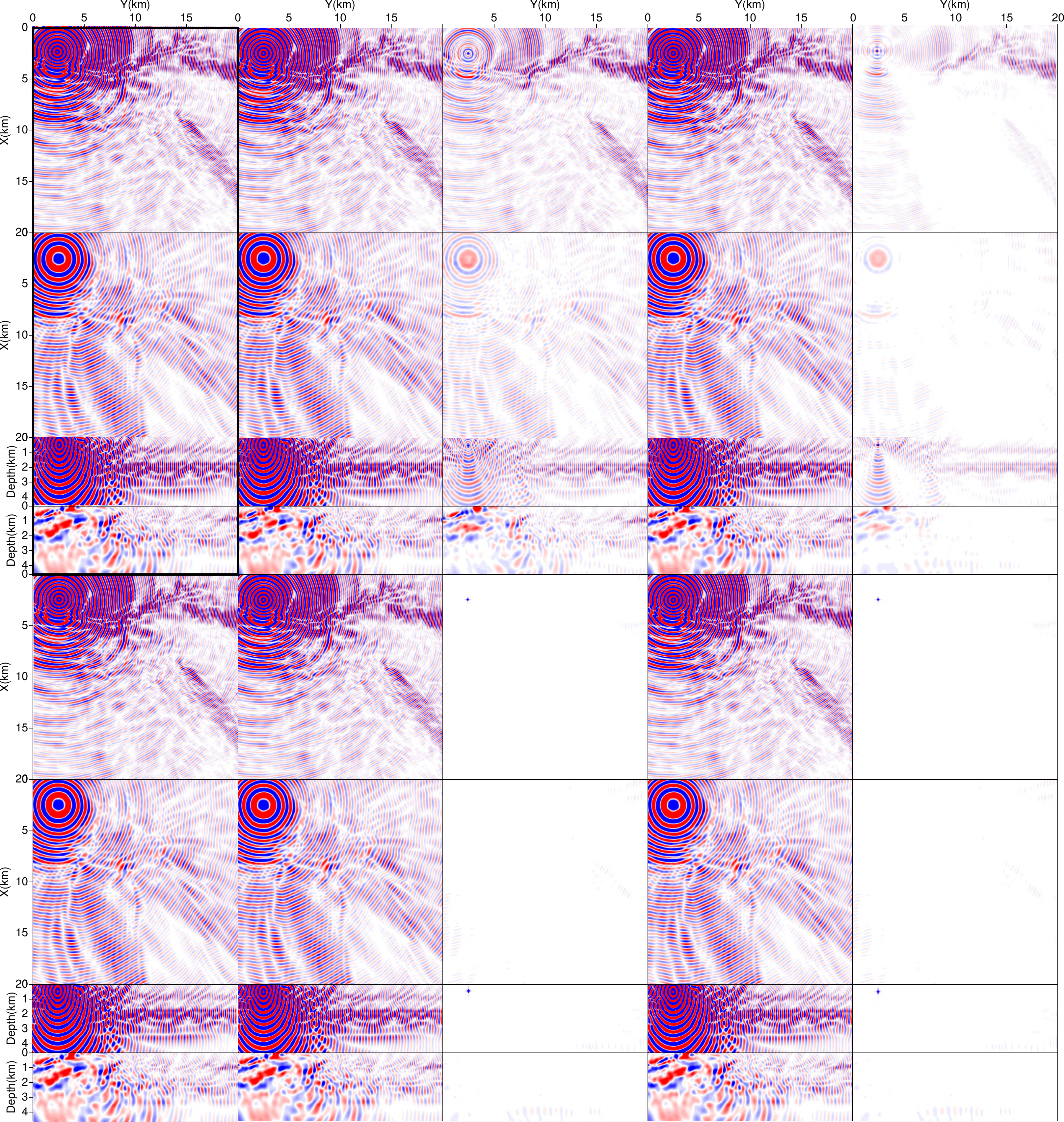}
\caption{Wavefields computed in the 3D SEG/EAGE Overthrust model. The real part if shown. The figure can be read as a $\bold{F}_{2 \times 5}$ matrix  where the first and second subscripts denote row and column, respectively. From top to bottom, one entry of the matrix (as an example, $\bold{F}_{1,1}$ is delineated by the thick line in the Figure) shows two depth sections at z=0.5~km depth and z=3.5~km depth and two vertical sections at y=2.5~km and y=15~km. ($\bold{F}_{1:2,1}$) CBS wavefield. ($\bold{F}_{1,2}$):  $G4$ wavefield. ($\bold{F}_{1,4}$) $Gm$ wavefield. ($\bold{F}_{2,2}$) $GA$ wavefield. ($\bold{F}_{2,4}$): $GAm$ wavefield. ($\bold{F}_{1,3}$, $\bold{F}_{1,5}$, $\bold{F}_{2,3}$, $\bold{F}_{2,5}$): Differences between CBS wavefield and $G4$, $Gm$, $GA$, $GAm$ wavefields.}
\label{fig_wavefield_overthrust}
\end{figure}

\begin{figure}
\center
\includegraphics[width=1\columnwidth]{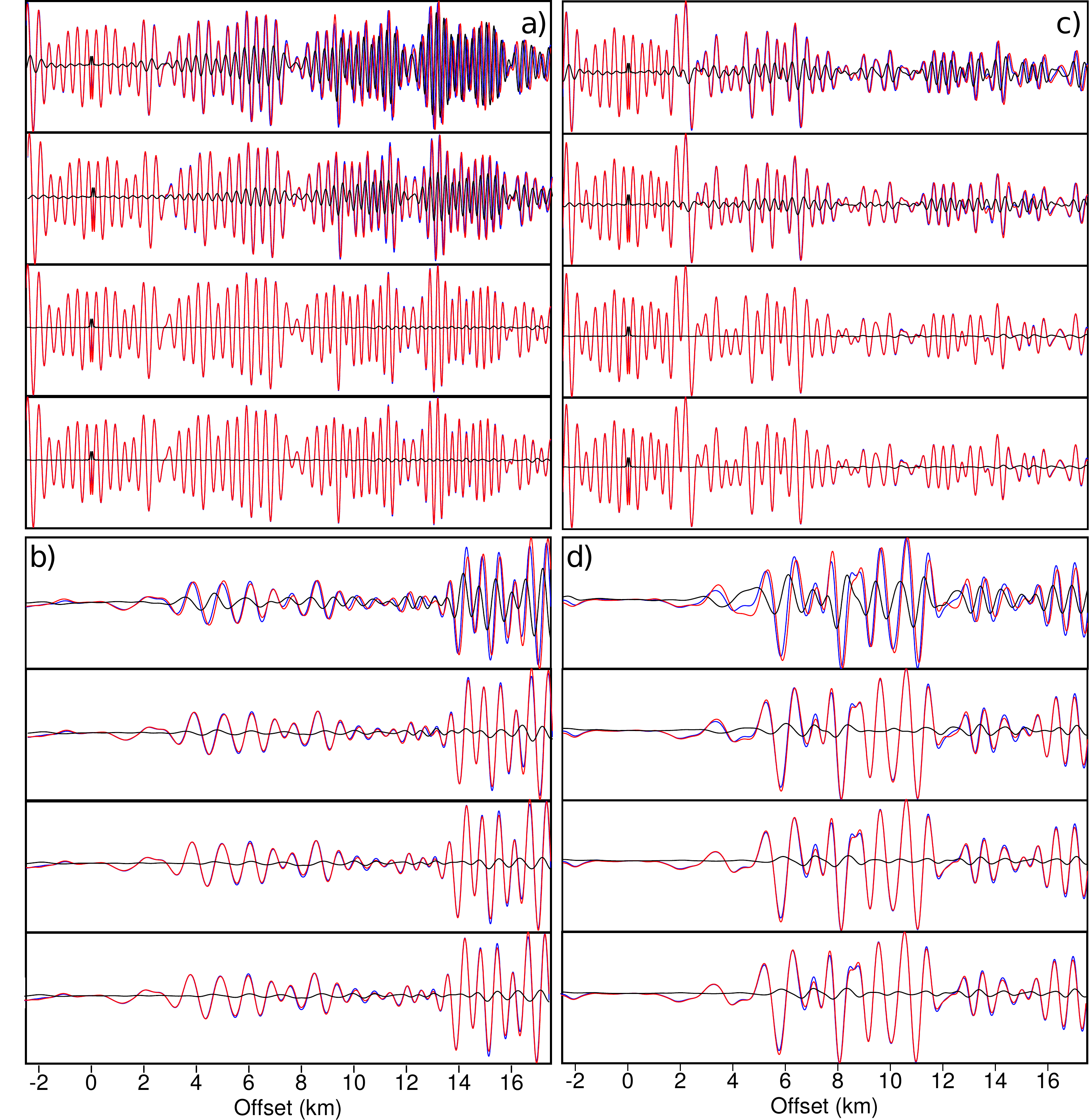}
\caption{3D SEG/EAGE Overthrust model. Direct comparison between horizontal profiles across the CBS (blue) and the FDFD (red) wavefields. The difference is plotted with black lines. Amplitudes are corrected from geometrical spreading through a linear gain with offset. In (a-d), from top to bottom, CBS versus $G4$, $Gm$, $GA$, and $GAm$ wavefields. (a) (x,z)=(2.5~km,0.5~km). (b) (x,z)=(15~km,3.5~km). (c) (a) (y,z)=(2.5~km,0.5~km). (b) (y,z)=(15~km,3.5~km)}
\label{fig_log_Overthrust}
\end{figure}

We now benchmark the adaptive stencil against more heterogeneous models involving either complex compressive tectonics (overthrust) or sharp contrasts (salt). We start with the overthrust model of dimensions 20~km $\times$ 20~km $\times$ 4.65~km. It represents a complex thrusted sedimentary succession on top of a structurally decoupled extensional and rift basement block \citep{Aminzadeh_1997_DSO} (Figure~\ref{fig_over_salt}a). A complex weathered zone also characterizes it in the near-surface with sharp lateral velocity variations and several sand channels. The velocities range between 2179~m/s at the surface and 6000~m/s at the basement. We re-mesh the model with a grid interval of 50~m and 8 grid points in the PMls along each face of the 3D grid of dimensions 417 $\times$ 417 $\times$ 109. The source is positioned at (2500~m,  2500~m, 500~m). The frequency is 10~Hz, and hence $G$ varies between ~4.4 and 12 (Table~\ref{tab_benchs}). We perform the simulation with 40 nodes of the Occigen computer. The elapsed time and the memory required to perform the LU factorization are 395~s and 1.2~Tera bytes. The wavefields are shown in Figures~\ref{fig_wavefield_overthrust} and \ref{fig_log_Overthrust}. In Figure~\ref{fig_wavefield_overthrust}, we show two depth slices of the wavefields at 500~m depth across the source and 3500~m depth just above the basement and two vertical sections at y=2500~m across the source and y=15000~m. In Figure~\ref{fig_log_Overthrust}, we show a direct comparison between the CBS and the FDFD wavefields along four horizontal profiles at (x,z)=(2000~m, 500~m), (x,z)=(15000~m, 3500~m), (y,z)=(2500~m, 500~m), (y,z)=(15000~m, 3500~m). As expected, the $Gm$ wavefield is more accurate than the $G4$ counterpart since the values of $G$ covered by the propagated wavelengths are consistent with the values of $G$ for which phase velocity dispersion was minimized during the estimation of the $Gm$ weights (Table~\ref{tab_accuracy}). The accuracy of the $GA$ and $GAm$ wavefields is significantly higher than that of the $Gm$ wavefield, hence providing a first validation of the adaptive stencil in heterogeneous media with the complex tectonic trend. The accuracies of the $GA$ and $GAm$ stencil are almost identical for this benchmark.
\subsection{3D SEG/EAGE Salt model}
\begin{figure}
\center
\includegraphics[width=1\columnwidth]{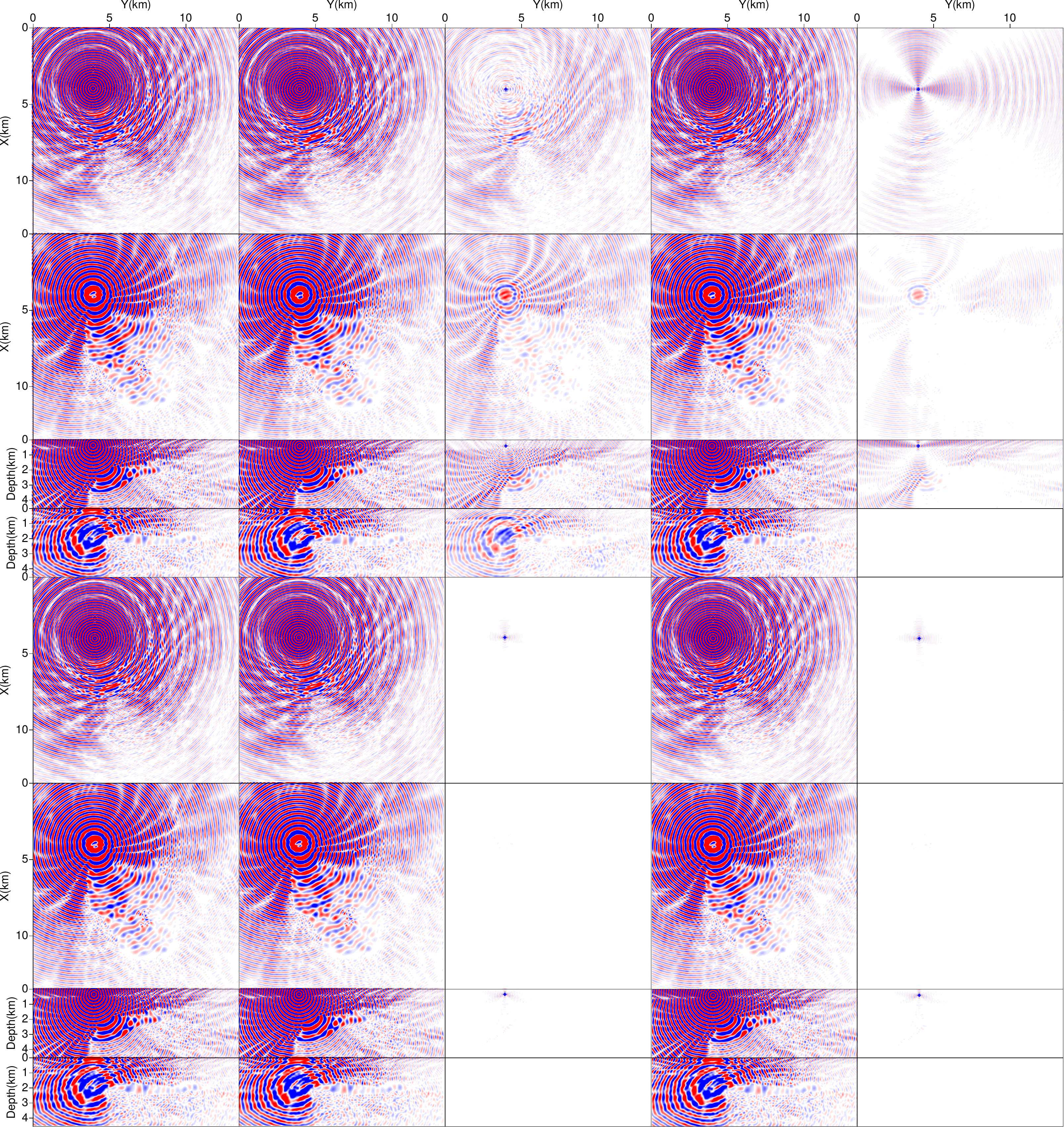}
\caption{Wavefields computed in the 3D SEG/EAGE Salt model. Same showing as that of Figure~\ref{fig_wavefield_overthrust} is used. The depth sections are at 0.4~km and 2~km depth and two vertical sections at y=4~km and y=8~km.}
\label{fig_wavefield_salt}
\end{figure}

\begin{figure}
\center
\includegraphics[width=1\columnwidth]{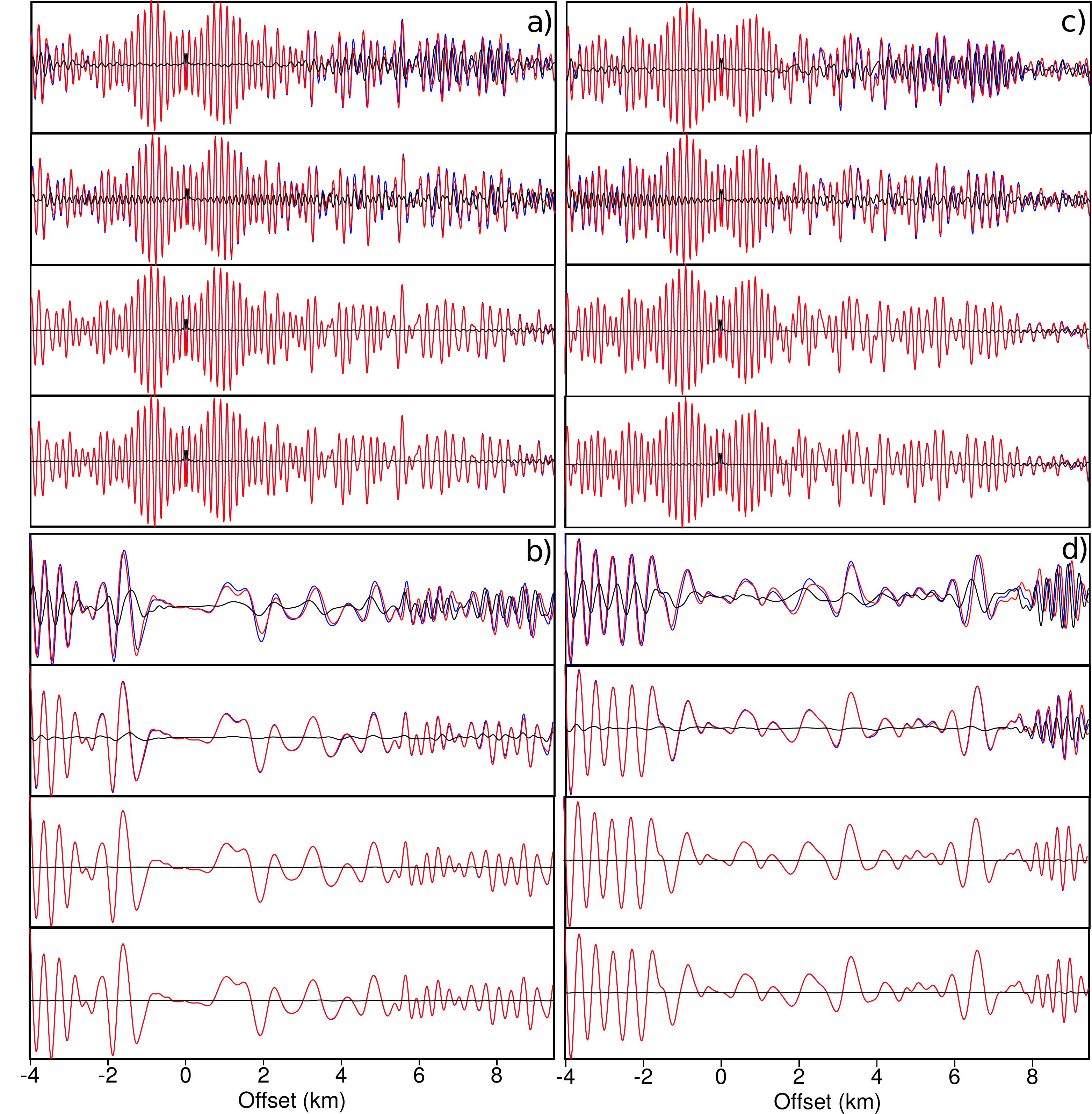}
\caption{Same as Figure~\ref{fig_log_Overthrust} for the Salt model. Positions of the horizontal profiles are (a) (x,z)=(4~km,0.4~km). (b) (x,z)=(8~km,,2~km). (c) (y,z)=(4~km,0.4~km). (d) (y,z)=(8~km,2~km).}
\label{fig_log_salt}
\end{figure}

\begin{figure}
\center
\includegraphics[width=1\columnwidth]{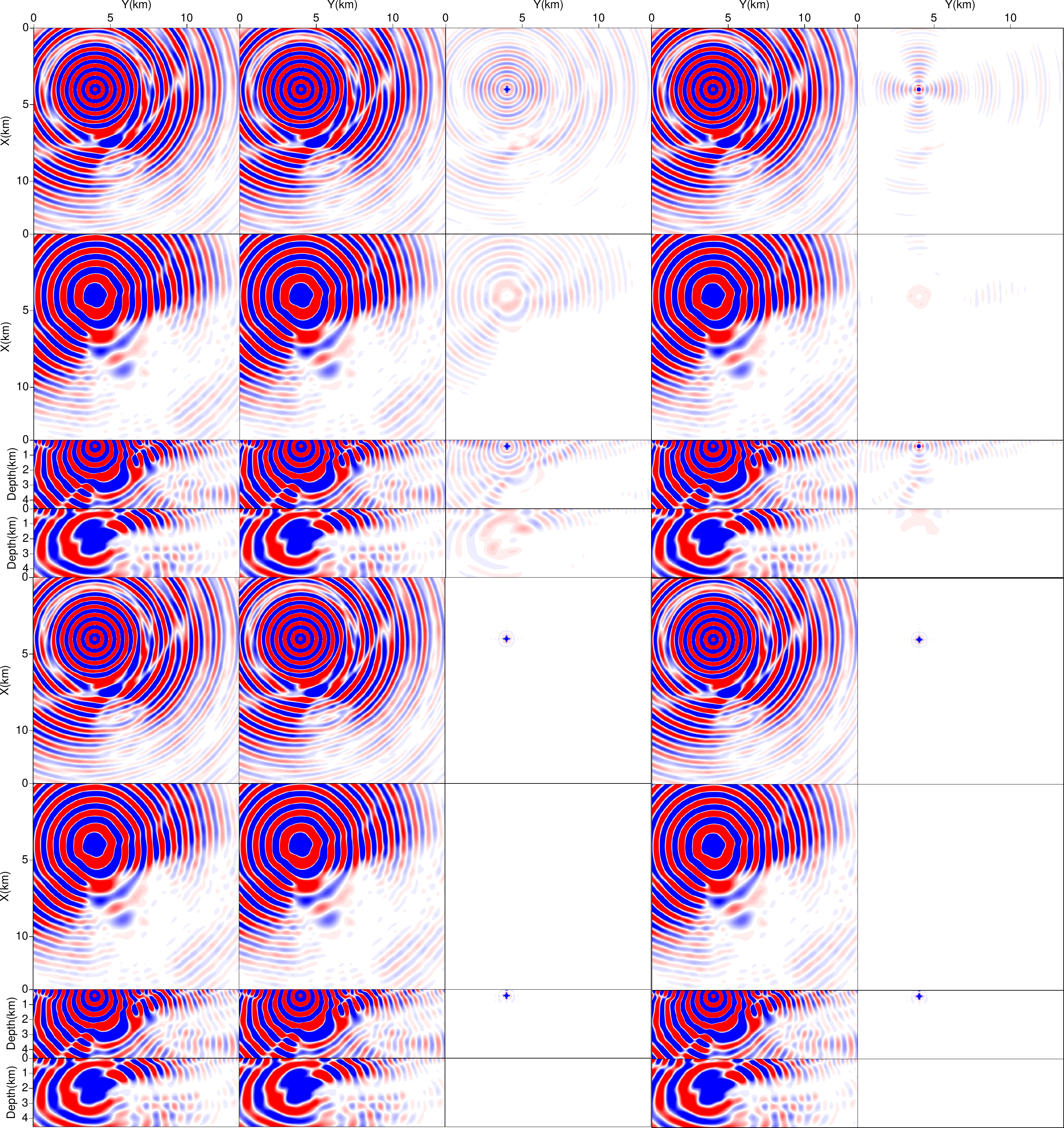}
\caption{Wavefields computed in the 3D SEG/EAGE Salt model. Same showing as that of Figure~\ref{fig_wavefield_salt} is used but frequency is 3~Hz.}
\label{fig_wavefield_salt_3Hz}
\end{figure}

\begin{figure}
\center
\includegraphics[width=1\columnwidth]{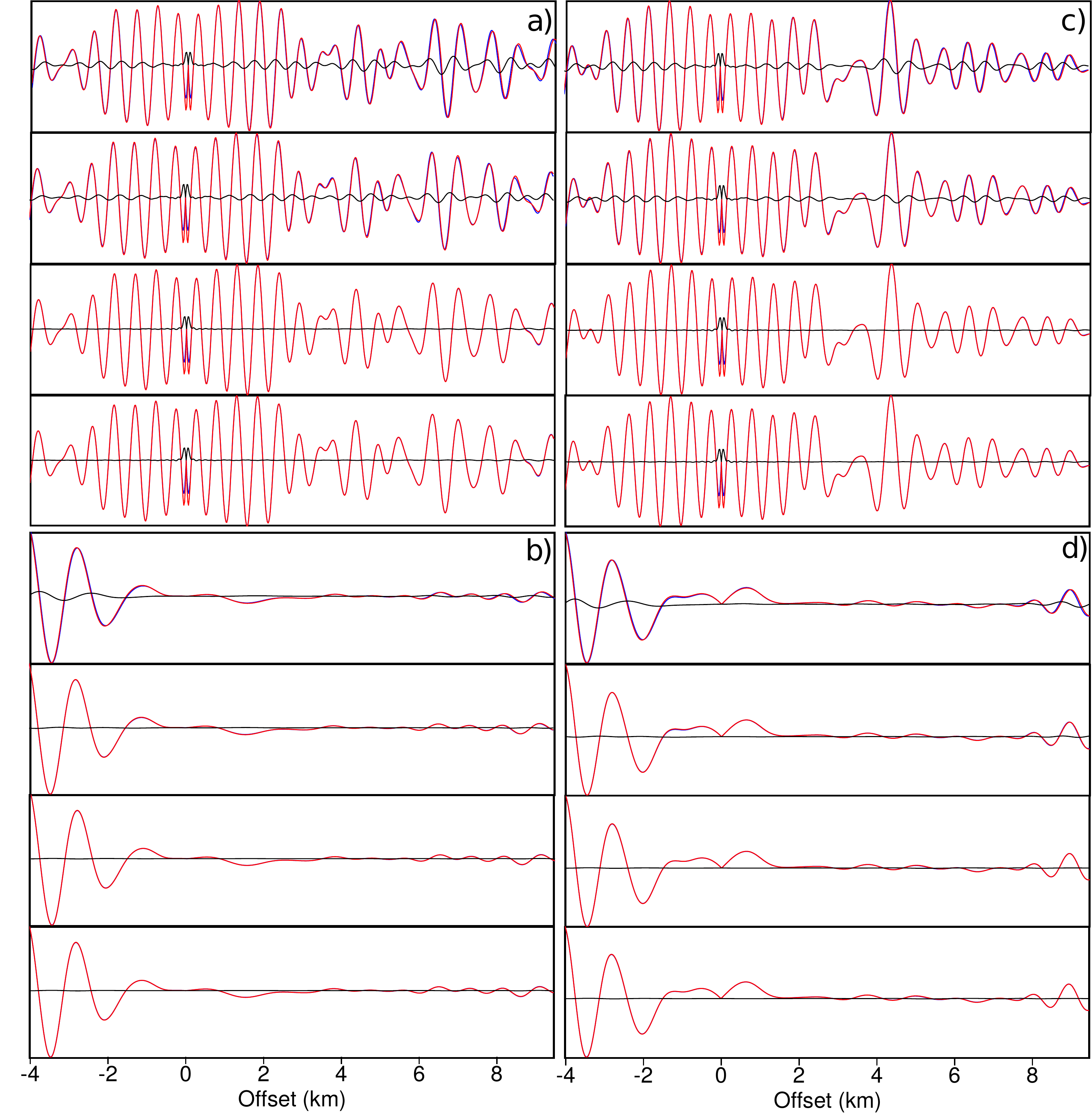}
\caption{Same as Figure~\ref{fig_log_salt} for the 3Hz frequency.}
\label{fig_log_salt_3Hz}
\end{figure}

We continue with the 3D SEG/EAGE salt model, which is representative of a typical Gulf coast salt structure \citep{Aminzadeh_1997_DSO} (Figure~\ref{fig_over_salt}b). Here, we want to check the ability of the adaptive stencil to deal with sharp velocity contrasts between the sediments and the salt. The velocities range between 1500~m/s in the water and 4481~m/s in the salt. We re-mesh the velocity model with a grid interval of 40~m and 8 grid points in the PMls along each face of the 3D grid of dimensions 354 $\times$ 354 $\times$ 131. The source is positioned at (4000~m, 4000~m, 400~m). The frequency is 9.375~Hz, and hence $G$ varies between 4 and 12 (Table~\ref{tab_benchs}). We perform the simulation with 40 nodes of the Occigen computer. The elapsed time and the memory required to perform the LU factorization are 448~s and 1.1~Tera bytes. The wavefields are shown in Figures~\ref{fig_wavefield_salt} and \ref{fig_log_salt}. 
Similar to the overthrust model, the $G4$ wavefield shows the worst accuracy. The $GA$ and $GAm$ wavefields show an impressive agreement with the CBS wavefield suggesting that the adaptive stencil is very valuable to deal with sharp contrasts.
This is highlighted by the fact that the relative accuracy improvement achieved by the adaptive stencils related to the $Gm$ counterpart is much more significant for the salt model compared to the overthrust model (Table~\ref{tab_accuracy}).
Unlike the overthrust model, the $GA$ scheme slightly outperforms the $GAm$ counterpart, further supporting the above statement. \\
We also check how the stencil behaves at lower frequencies. We re-mesh the model with a grid interval of 80~m and perform the simulation at the 3~Hz frequency leading to $G$ ranging between 6.25 and 18.7. Again the adaptive stencils achieve an impressive accuracy (Figures~\ref{fig_wavefield_salt_3Hz} and \ref{fig_log_salt_3Hz}).
\subsection{3D GO\_3D\_OBS crustal model}
\begin{figure}
\center
\includegraphics[width=1\columnwidth]{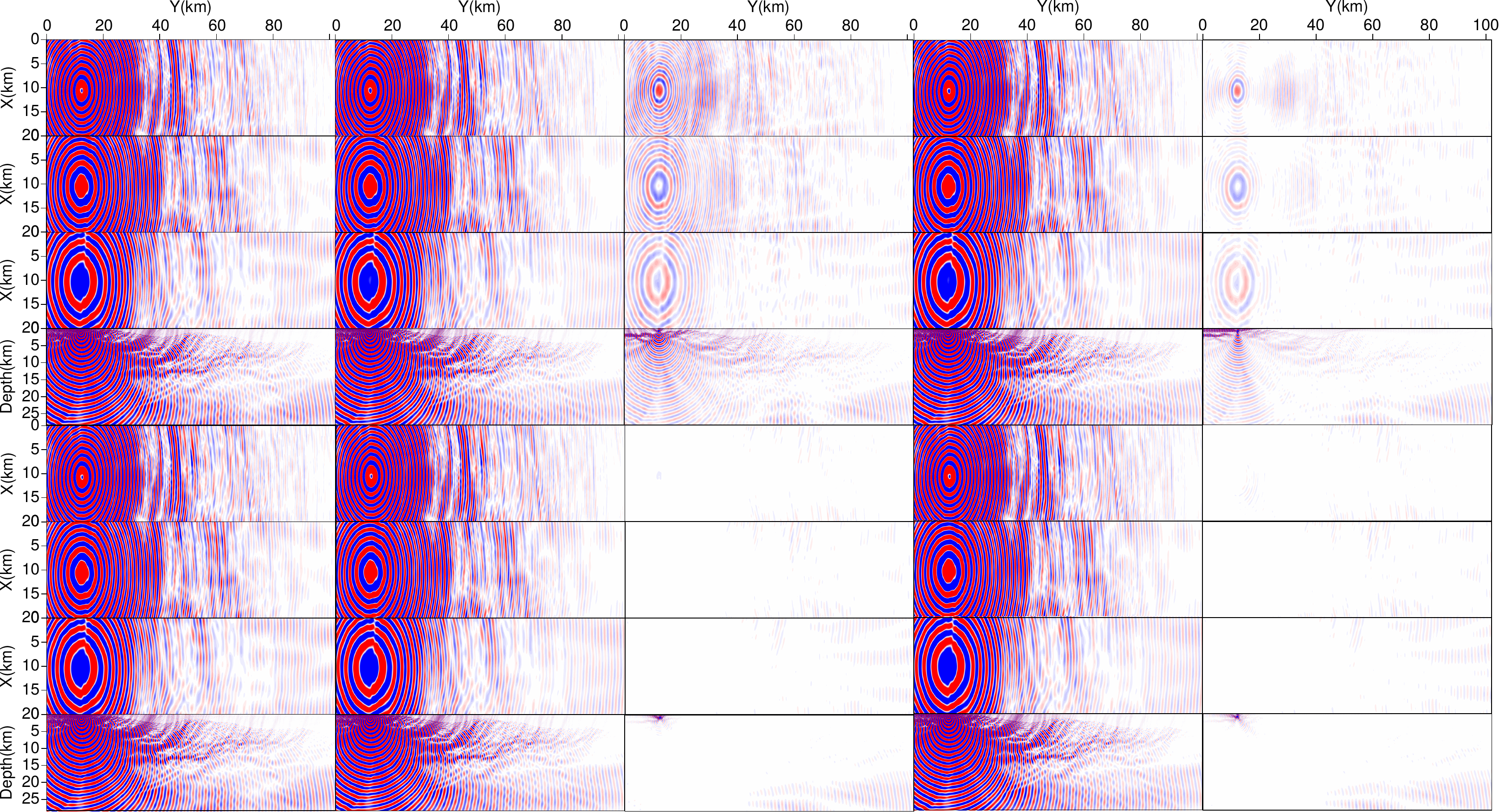}
\caption{Wavefields computed in the 3D GO\_3D\_OBS model. Same showing as that of Figure~\ref{fig_wavefield_overthrust}.}
\label{fig_wavefield_go3dobs}
\end{figure}

\begin{figure}
\center
\includegraphics[width=1\columnwidth]{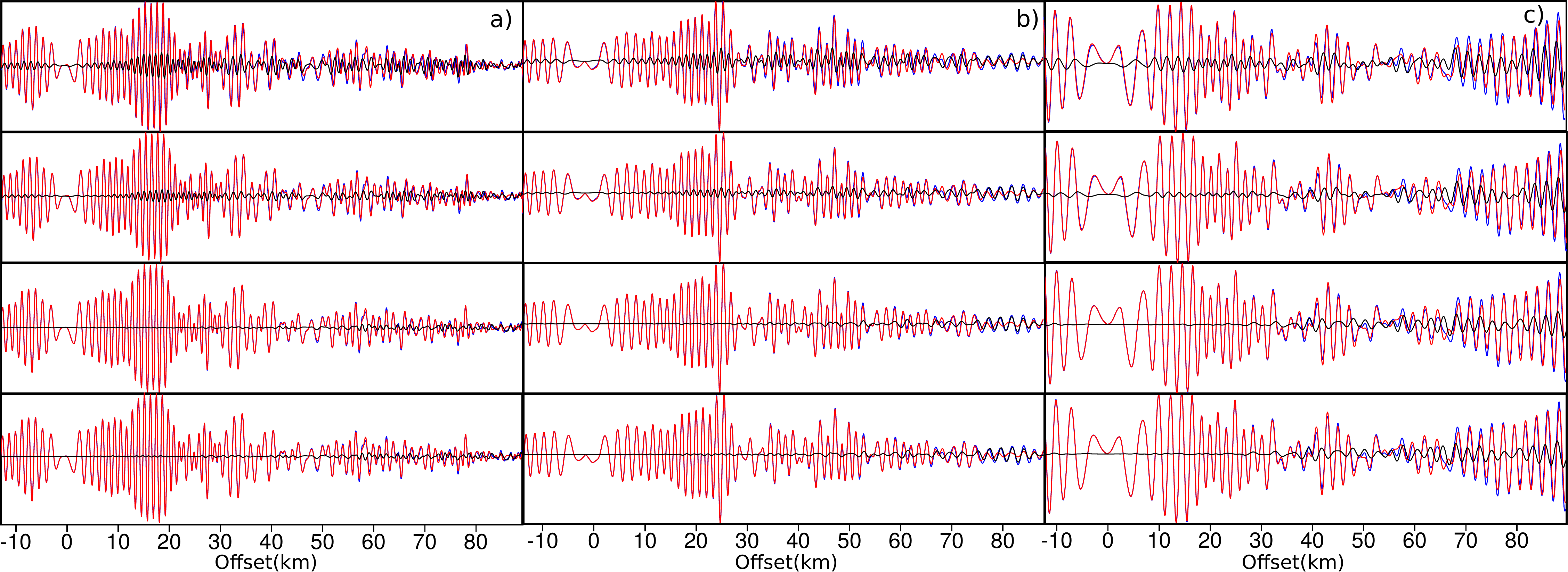}
\caption{Direct comparison between y-profiles of the real part of the CBS (blue) and the FDFD (red) wavefields across the source position (x=10.5~km). The differences are plotted in black. Amplitudes are corrected for geometrical spreading. Depth of profiles are 6~km, 10~km and 16~km.}
\label{fig_log_go3dobs}
\end{figure}

The last benchmark corresponds to the 3D GO\_3D\_OBS crustal model \citep{Gorszczyk_2021_GNT}. This geomodel has been designed to assess seismic imaging techniques for regional deep crustal exploration. It embeds the main structural factors that characterize subduction zones and has been inspired by the FWI case study performed in the eastern Nankai trough by \citet{Gorszczyk_2017_TRW}. We select a target of the model of dimensions 20~km $\times$ 102~km $\times$ 28.4~km. We discretize this target with a grid interval of 100~m with eight grid points in the PMLs, leading to a finite-difference grid of dimensions 217 $\times$ 1037 $\times$ 300, hence 67.5~millions degrees of freedom. We show the result of the simulation for a  source positioned on the sea bottom at (10500~m, 12300~m, 900~m). The simulated frequency is 3.75~Hz. The wavespeeds range between 1500~m/s in the water and 8639.1~m/s in the upper mantle, and hence $G$ ranges between 4 and 23 (Table~\ref{tab_benchs}). We perform the simulation using 120 nodes of Occigen computer. The elapsed time for factorization and the elapsed time to compute 130 wavefield solutions associated with a sparse layout of ocean bottom nodes is 2615~s and 50~s, respectively. The used peak memory is 9.17~Tera bytes. Note that much improved performances can be obtained in such kinds of large computational domains with the block-low rank version of MUMPS \citep{Amestoy_2016_CPB,Amestoy_2016_FFF,Amestoy_2018_CBL}. This topic will be addressed in the next study. Note also that this simulation could have been performed at the 5~Hz frequency with good accuracy. For this frequency, $G$=3 in the water, which is undersampled according to the FWI resolution. However, the water layer can be kept fixed during FWI since the wavespeed in water is know with good accuracy. Comparisons between the CBS and FDFD wavefields are shown in Figures~\ref{fig_wavefield_go3dobs} and \ref{fig_log_go3dobs}. In Figure~\ref{fig_log_go3dobs}, a direct comparison between the CBS and wavefields is performed along with three horizontal profiles in the $y$ direction running across the source position and cross-cutting the accretionary wedge at 6~km depth, the subduction megathrust at 10~km depth and the Moho at 16~km depth. The $G4$ wavefield lacks accuracy since this value of $G$ is met in the water only. The $GA$ and $GAm$ wavefields are significantly more accurate than the $Gm$ counterpart. The $GA$ wavefield is slightly more accurate than the $GAm$ counterpart, further supporting that the adaptive stencil doesn't require averaging of the weights. Note that compared to the previous tests, we see more significant errors in the bottom-right part of the computational domain for the $Gm$, $GA$, and $DAm$ wavefields. These errors may result from the higher number of propagated wavelengths in the GO\_3D\_OBS crustal model. They may also result from some inaccuracies of the PMLs at grazing incidences. 
%

\subsection{Comparison with high-order finite-difference time-domain method}
\begin{figure}
\center
\includegraphics[width=1\columnwidth]{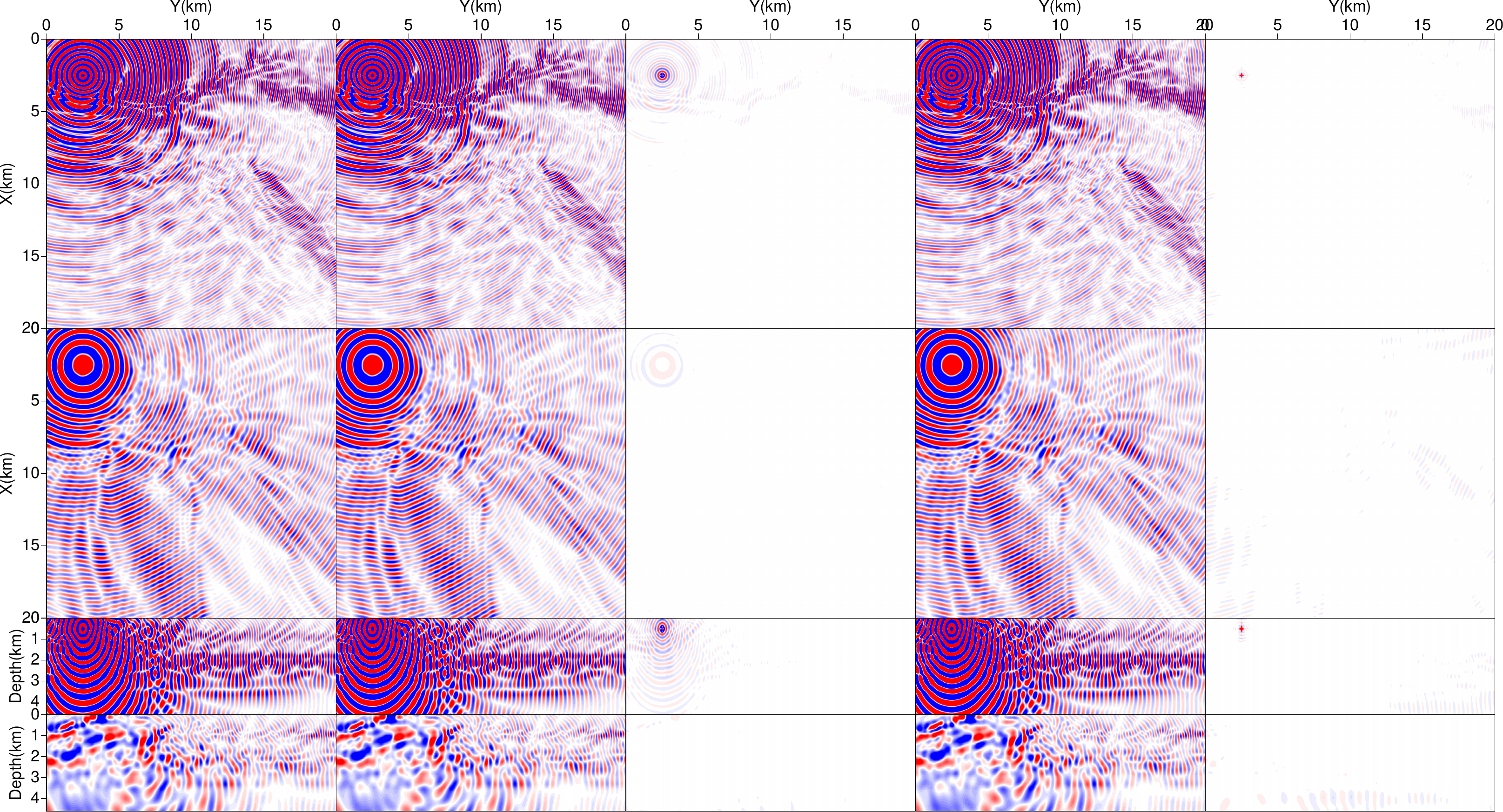}
\caption{Wavefields computed in the 3D SEG/EAGE Overthrust model. The real part is shown. The figure can be read as a $\bold{F}_{1 \times 5}$ matrix  where the first and second subscript denote row and column, respectively. From top to bottom, one entry of the matrix shows two depth sections at z=0.5~km and z=3.5~km and two vertical sections at x=2.5~km and x=15~km . ($\bold{F}_{1,1}$) CBS. ($\bold{F}_{1,2}$):  $\mathcal{O}(\Delta t^2,\Delta x^8)$ FDTD solution. ($\bold{F}_{1,4}$) FDFD-Ga. ($\bold{F}_{1,3}$, $\bold{F}_{1,5}$): Differences between CBS and FDTD, FDFD-Ga.}
\label{fig_wavefield_overthrust_fdtd}
\end{figure}

\begin{figure}
\center
\includegraphics[width=1\columnwidth]{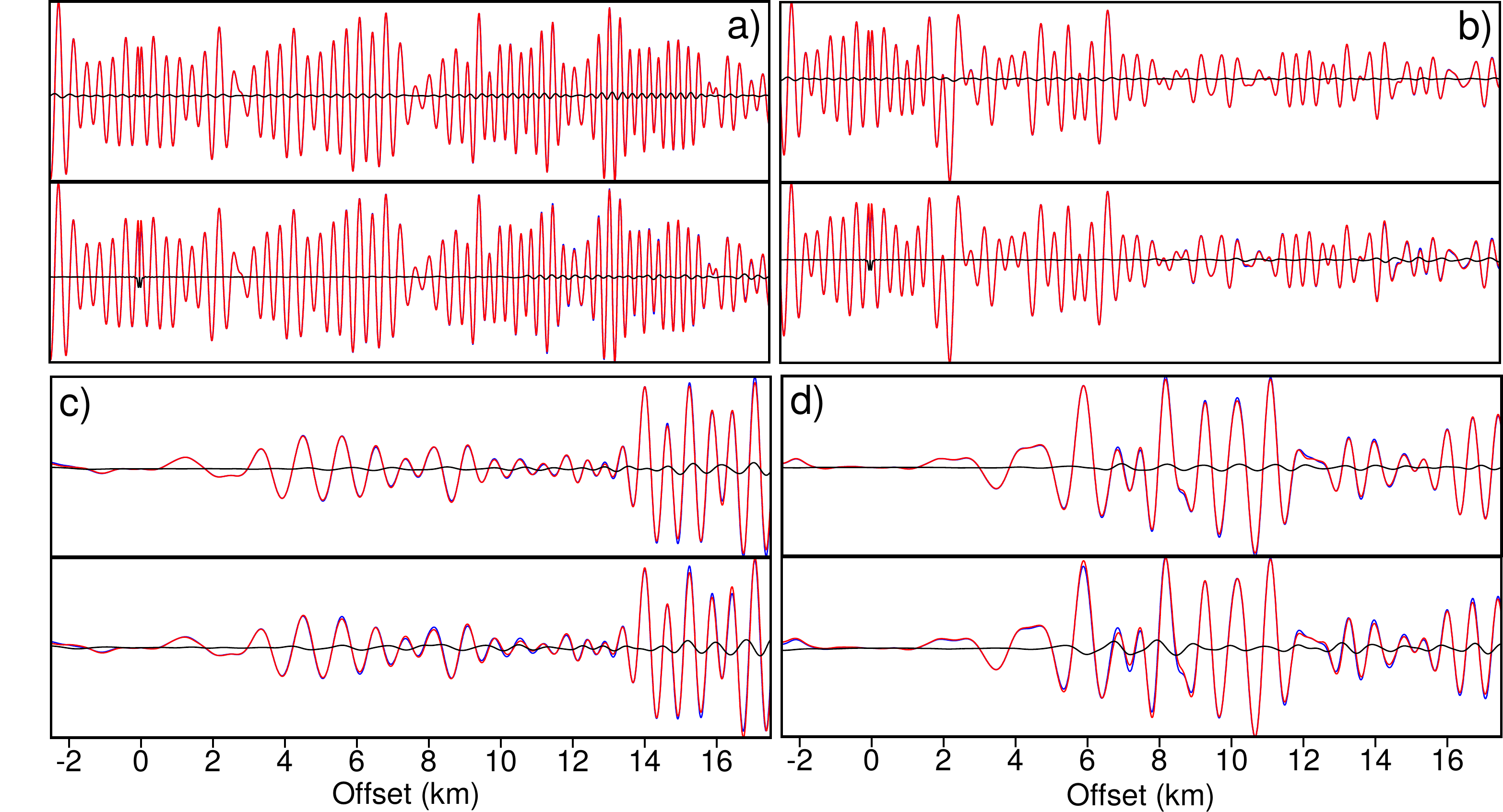}
\caption{3D SEG/EAGE Overthrust model. Direct comparison between horizontal profiles across the CBS (blue), FDTD and FDFD (red) wavefield solutions. The difference is plotted with black lines. Amplitudes are corrected from geometrical spreading through a linear gain with offset. In (a-d), from top to bottom, CBS versus FDTD and CBS versus FDFD-Ga. (a) (x,z)=(2.5~km,0.5~km). (b) (x,z)=(15~km,3.5~km). (c) (a) (y,z)=(2.5~km,0.5~km). (b) (y,z)=(15~km,3.5~km).}
\label{fig_log_Overthrust_fdtd}
\end{figure}

\begin{figure}
\center
\includegraphics[width=1\columnwidth]{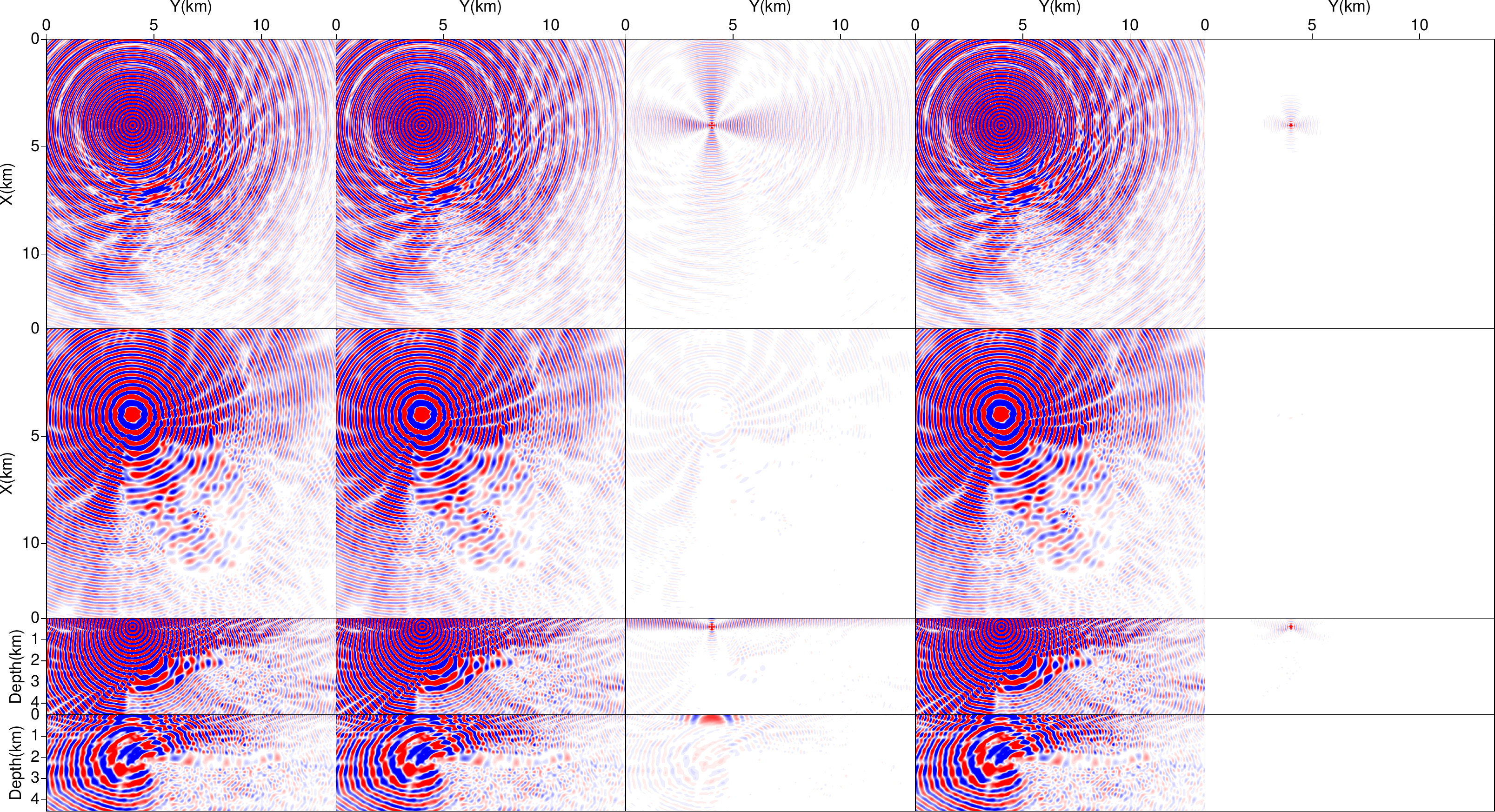}
\caption{Wavefields computed in the 3D SEG/EAGE salt model. The real part is shown. Same as Figure~\ref{fig_wavefield_overthrust_fdtd} but for the salt model.}
\label{fig_wavefield_salt_fdtd}
\end{figure}

\begin{figure}
\center
\includegraphics[width=1\columnwidth]{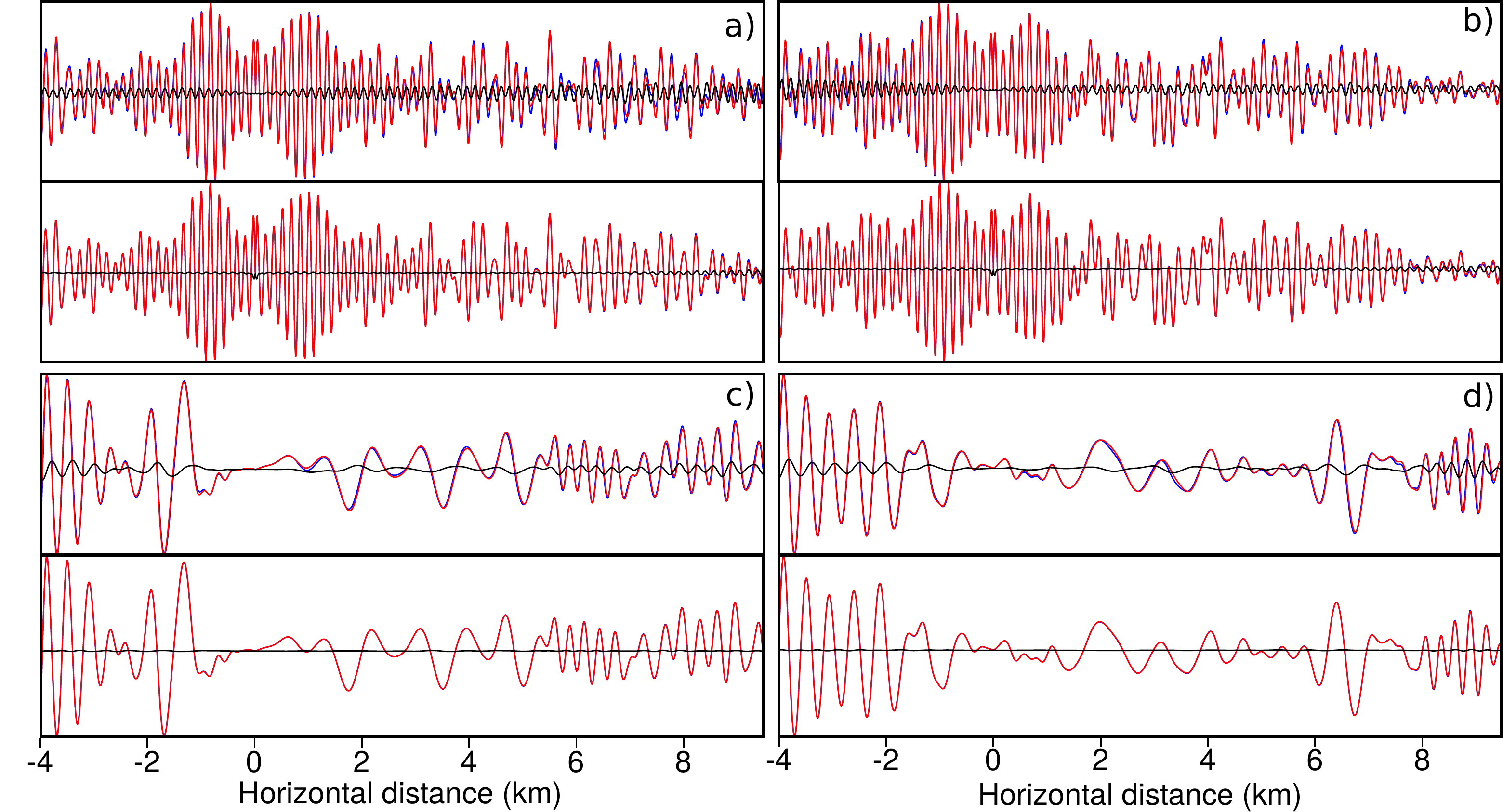}
\caption{3D SEG/EAGE salt model. Direct comparison between horizontal profiles across the CBS (blue), FDTD and FDFD (red) wavefield solutions. The difference is plotted with black lines. Amplitudes are corrected from geometrical spreading through a linear gain with offset. In (a-d), from top to bottom, CBS versus FDTD and CBS versus FDFD-Ga. (a) (x,z)=(4~km,0.4~km). (b) (x,z)=(8~km,2~km). (c) (a) (y,z)=(4~km,0.4~km). (b) (y,z)=(8~km,2~km).}
\label{fig_log_salt_fdtd}
\end{figure}

It is also worth benchmarking the adaptive FDFD method against a high-order staggered-grid velocity-stress finite-difference time-domain method (FDTD) \citep{Virieux_1984_SWP,Levander_1988_FOF,Graves_1996_SSW}. In the latter case, the monochromatic solution is computed on the fly in the loop over time steps by discrete Fourier transform \citep{Nihei_2007_FRM,Sirgue_2008_FDW}. The FDTD stencil is second-order accurate in time and eight-order accurate in space. We use a Ricker wavelet for these simulations instead of a delta signature to prevent numerical artifacts during the FDTD simulations. We first compare the $CBS$ wavefield, the $GA$ wavefield and the monochromatic wavefield inferred from the FDTD simulation for the overthrust model (Figures~\ref{fig_wavefield_overthrust_fdtd}-\ref{fig_log_Overthrust_fdtd} and Table~\ref{tab_accuracy1}). The wavefields computed with the two finite-difference methods match pretty well the CBS wavefield. Using the global metric given by equation~\ref{eqerror}, the FDTD solution is in average more accurate than the FDFD counterpart. However, a close examination of the profiles in Figure~\ref{fig_log_Overthrust_fdtd} shows that the FDFD wavefield is clearly more accurate than the FDTD counterpart in the shallow part, while the FDTD wavefield is more accurate in the deep part. This results because the accuracy of the FDTD and FDFD methods is managed with different paradigms:  the error increases monotonically with $1/G$ from 0 at the zero frequency to higher values near the Nyquist frequency in FDTD where the derivative is approximated by a Taylor polynomial of a given degree \citep{Brac_1997_SPO}, while the accuracy of the adaptive FDFD is optimized for each value of $G$ through the weight estimation. Accordingly, the error doesn't necessarily increase monotonically with $1/G$ in FDFD, as illustrated for example in Figure~\ref{fig_dispersion}(a-b). This explains why the FDFD wavefields are more accurate for small values of $G$, while high-order FDTD is more accurate for high values of $G$ as the error tends to 0. \\
We repeat this comparison for the salt model. The results are shown in Figures~\ref{fig_wavefield_salt_fdtd} and \ref{fig_log_salt_fdtd} and in Table~\ref{tab_accuracy1}. Interestingly, the $GA$ wavefield is significantly more accurate than the FDTD counterpart for this benchmark involving sharp contrasts. This highlights how large-support high-order stencils implemented with high-degree Taylor polynomials implemented in FDTD tend to smooth out the effects of the contrasts, unlike the compact-support adaptive FDFD stencil. 

\begin{table}
\begin{center}
\caption{Main specifications of the five benchmark models. $dof$: Number of degrees of freedom in the FD grid (including PMls). $c_{m}$, $c_{M}$: Minimum and maximum wavespeeds. $f$: Frequency. $h$: Grid interval. $\lambda_{min}$,  $\lambda_{max}$: Minimum and maximum wavelength. $G_{min}$, $G_{max}$: Smallest and higher $G$. $N_\lambda$: Maximum number of propagated wavelengths.}
\label{tab_benchs}
\begin{tabular}{|c|c|c|c|c|c|c|c|c|c|c|}
\hline
Models      & ndof(10$^6$)    &  $c_{m}(m/s)$ 	&  $c_{M}(m/s)$	& $f(Hz)$ 	& $h(m)$ & $\lambda_{m}(m)$ & $\lambda_{M}(m)$ & $G_{m}$ & $G_{M}$ & $N_\lambda$		\\ \hline
Homogeneous & 23.5	& 1500 & 1500 & 7.5 & 50 & 375	& 375 & 4 &  4	&  40	\\ \hline
Linear 		& 23.5 &  1500 & 8500 & 7.5 & 50 & 375	& 1133 & 4 & 22.7 &	29	\\ \hline
Overthrust 	&  17.2 & 2179 & 6000 & 10 & 50 & 	218 & 600 & 4.4 & 12 &	50		\\ \hline
Salt  		 &	16.4 & 1500 & 4482 & 9.375 & 40 & 160	& 478 & 4 & 12	&	48	\\ \hline
GO\_3D\_OBS 	& 67.5 & 1500 & 8639.1 & 3.75 & 100 & 400	& 2303.8 & 4 & 23 &	255		\\ \hline
\end{tabular}
\end{center}
\end{table}
\begin{table}
\caption{Accuracy measurement of the 27-point FDFD stencil for the different benchmarks. For the homogeneous and linear models, the reference solutions are the analytical ones. For the other models, the reference solution is the CBS one. The errors between the reference wavefield and the $G4$, $Gm$, $GA$ and $GAm$ wavefields(equation~\ref{eqerror}) are shown in the table.}\label{tab_accuracy}
\begin{center}
\begin{tabular}{|c|c|c|c|c|}
\hline
Models          &  G4 	& Gm 	& GA 	& GAm	\\ \hline
Homogeneous 	&  0.0242  &   0.8065 & 0.0252	&    0.0252			\\ \hline
Linear 		&  0.4822  &  0.2285  &  0.013	&  0.0135 		\\ \hline
Overthrust 	&  0.8619  &  0.3123  &  0.129	&   0.1289  			\\ \hline
Salt  		&  0.7689 &  0.2582  &  0.0298 &    0.031		\\ \hline
GO\_3D\_OBS 	&  0.4573  & 0.3022   &  0.2114	&  0.2125   			\\ \hline
\end{tabular}
\end{center}
\end{table}
\begin{table}
\caption{Accuracy of the $GA$ stencil versus the $\mathcal{O}(\Delta t^2,\Delta h^8)$ staggered-grid finite-difference time-domain stencil for the overthrust and salt benchmarks. The error given by equation~\ref{eqerror} is provided in the table.}
\label{tab_accuracy1}
\begin{center}
\begin{tabular}{|c|c|c|}
\hline
Models          &  FDTD error  	& FDFD-GA error	\\ \hline
Overthrust 	&  0.0876   &  0.1278   			\\ \hline
Salt  		&  0.1977 &     0.0297			\\ \hline
\end{tabular}
\end{center}
\end{table}
\section{Conclusion}  
We significantly improve the accuracy of the 27-point finite-difference stencil for 3D frequency-domain (visco-)acoustic seismic wave modeling by adapting the weights through which dispersion is minimized to the local wavelength.
This wavelength-adaptivity doesn't introduce computational overhead because the shape (position and number of coefficients) of the stencil is unchanged, and the weights are tabulated once and for all for a wide range of the number of grid points per wavelength. Therefore, any impedance matrix associated with a particular simulation is easily built by picking the appropriate weights in the table for each of its rows. We benchmark the method with different subsurface models of increasing complexity. In all cases, the adaptive stencil outperforms the non-adaptive stencil in terms of accuracy. The accuracy improvement is more significant when the subsurface model contains sharp contrasts. The compact adaptive stencil also shows improved accuracy compared to the high-order finite-difference time-domain method in the presence of sharp contrasts. The method should also apply to acoustic media with transverse isotropic effects and elastic media. 
\section*{Acknowledgments}  
This study was partially funded by the WIND consortium (\textit{https://www.geoazur.fr/WIND}), sponsored by Chevron, Shell and Total. The authors are grateful to the OPAL infrastructure from 
Observatoire de la Côte d'Azur (CRIMSON) for providing resources and support. This work was granted access to the HPC resources of IDRIS under the allocation A50050410596 made by GENCI. We thank the MUMPS developers, P. Amestoy (Mumps Technologies - ENS Lyon), A. Buttari (CNRS-IRIT), J-Y. L'Excellent (Mumps Technologies - ENS Lyon), T. Mary (LPI-CNRS), C. Puglisi (Mumps Technologies - ENS Lyon), for providing the MUMPS solver (\url{http://mumps-solver.org}) and assistance for optimal tuning.

\bibliographystyle{gji}
\newcommand{\SortNoop}[1]{}

\end{document}